\documentstyle{article}
\def\openone{\leavevmode\hbox{\small1\kern-3.8pt\normalsize1}}

\newcommand{\newsection}[1]{\setcounter{subsection}{0}
			\vspace{4ex} \addtocounter{section}{1} \noindent
			{\Large \bf \Roman{section}. #1} \vspace{3ex} }

\begin{document}

\mbox{}
\vspace{10ex}
\begin{center}
{\huge \bf Canonical Transformations and} \\
\mbox{} \\
{\huge \bf  Path Integral Measures} \\
\mbox{} \\
\mbox{} \\
Mark S.~Swanson\footnote{E--Mail Address: {\bf Swanson@UConnvm.UConn.Edu} } \\
Department of Physics \\
University of Connecticut \\
Stamford, Connecticut 06903
\end{center}

\vspace{6ex}
\begin{center}
Abstract
\end{center}

\begin{quote}

This paper is a generalization of previous work on the use of classical
canonical transformations to evaluate Hamiltonian path integrals for quantum
mechanical systems.  Relevant aspects of the Hamiltonian path integral and
its measure are discussed and used to show that the quantum mechanical
version of the classical transformation does not leave the measure of the
path integral invariant, instead inducing an anomaly.  The relation to
operator techniques and ordering problems is discussed, and special
attention is paid to incorporation of the initial and final states of the
transition element into the boundary conditions of the problem.  Classical
canonical transformations are developed to render an arbitrary power
potential cyclic.  The resulting Hamiltonian is analyzed as a quantum system
to show its relation to known quantum mechanical results.  A perturbative
argument is used to suppress ordering related terms in the transformed
Hamiltonian in the event that the classical canonical transformation leads
to a nonquadratic cyclic Hamiltonian.  The associated anomalies are analyzed
to yield general methods to evaluate the path integral's prefactor for such
systems.  The methods are applied to several systems, including linear and
quadratic potentials, the velocity-dependent potential, and the
time-dependent harmonic oscillator.

\end{quote}

\vspace{30ex}

\pagebreak

\newsection{Introduction}

The relationship between classical mechanics and its quantum counterpart is
nowhere more evident than in the path integral formulation of transition
amplitudes\cite{Feyn,Schul,Swan1}, where the classical action evaluated
along a possible trajectory appears as a weighting phase factor for the
trajectory.  In the Hamiltonian form for the path integral the classical
action appears, at least formally, written in terms of the canonically
conjugate variables $q$ and $p$.  In classical mechanics it is precisely
this form of the action that is used to define canonical transformations
\cite{Gold} to new canonical variables, {\it i.e.}, those that preserve the
Poisson bracket structure. With an appropriate choice of canonical
transformation the classical action can be transformed to cyclic coordinates
or, in the case of the Hamilton--Jacobi equation, the Hamiltonian can be
transformed identically to zero.  At the classical level the solution of
Hamilton's equations becomes trivial for such a Hamiltonian.

For certain transition elements the measure of the corresponding Hamiltonian
path integral is symmetric in its integrations over intermediate $p$ and $q$
values and is therefore invariant under a transformation that has a Jacobian
of unity.  Since the Poisson bracket of a canonical transformation is
identical to the inverse Jacobian of the transformation, a canonical
transformation apparently introduces no factors into the measure.  At first
glance it would then seem that the classical canonical transformation to
cyclic variables could be applied to the Hamiltonian path integral to render
it exactly integrable, thus providing a method for nonperturbative analysis
of transition elements. Further analysis reveals that the transformed path
integral yields results differing from those of the original untransformed
path integral, and indeed those of other methods such as wave mechanics. In
particular, the time-dependent prefactor or, loosely speaking, the van Vleck
determinant\cite{Vleck} is incorrectly calculated using this method.  The
loss of such information is catastrophic to understanding the thermal
behavior, the stability, and tunneling rates of the system.  Perhaps more
disturbing, the invariance of the Hamiltonian path integral measure under
canonical transformations is a central assumption in the path integral
method for implementing first and second class constraints \cite{Fad,Senj},
a method whose generalization to gauge theories was seminal to their
quantization.  Recently developed equivariant localization techniques
\cite{Niemi} also rely on the ability to transform the Hamiltonian path
integral to new coordinates while introducing no unexpected terms in the
measure.

The problem, as discussed by numerous authors
\cite{Edwards,Schul2,Gervais,Fanelli,Prok,Gerry}, lies in the action
appearing in the path integral.  The time-derivatives appearing in the path
integral action are {\em formal}\/ identifications only, behaving like the
derivative familiar from calculus only for certain systems.  As a result,
the classical canonical transformation does not have the same result when it
is applied to the path integral action, and in fact it induces additional
terms into the action.  An alternative approach to canonical transformations
for the path integral is to define a quantum mechanical version of the
canonical transformation that is consistent with the formal time-derivatives
of the path integral\cite{Fukutaka,Swan2}.  Such an approach will be
followed in this paper.  However, this quantum canonical transformation
neither leaves the measure invariant, instead inducing nontrivial Jacobians,
nor necessarily reproduces the classical result for the transformed
Hamiltonian.  In a previous paper that concentrated on the measure
\cite{Swan2} it was shown that the induced Jacobians could be absorbed into
the action of the path integral where they appear as $O(\hbar)$ terms.
Alternatively, the Jacobian was shown to be equivalent to a time-dependent
prefactor that reproduced the van Vleck determinant of the original path
integral, at least for the case of the simple harmonic oscillator.  This
paper will consider both aspects of quantum canonical transformations and
generalize the previous results, fleshing out derivations along the way.  In
addition, new results regarding the application of canonical transformations
to the path integral as well as relevant properties of the path integral
will be presented and demonstrated for various systems.

The outline of this paper is as follows.  In Sec.~II the quantum mechanical
transition element to be analyzed will be expressed as a Hamiltonian path
integral.  Certain properties of expectation values necessary to the
remainder of the paper will be derived from this path integral for general
cases of the Hamiltonian.  The relation of the initial and final states to
the boundary values of Hamilton's equations are discussed, since this will
be of importance to the quantum canonical transformation.  It is shown that
there exist classically suppressed potentials, {\it i.e.}, ones that would not
contribute to the classical variation of the action, that contribute $O (
\hbar )$ terms to the quantum mechanical action.  Poisson resummation
techniques are applied to path integrals with periodic boundary conditions,
such as the square well system, to transform the measure to continuous
variables. The relation of continuum techniques to the boundary conditions
are discussed for the Hamiltonian path integral.  In Sec.~III the classical
canonical transformation will be reviewed with special attention to the
relation of the surface or endpoint terms generated by the canonical
transformation to the boundary conditions of the transition element.  The
operator ordering ambiguities associated with the quantum mechanical version
of these transformations are briefly discussed and the specific problem of
Cartesian versus polar coordinates is used to demonstrate them.  A classical
canonical transformation to cyclic coordinates is derived for the case of an
arbitrary power potential. The resulting classical Hamiltonian is analyzed
as a quantum mechanical problem, thereby ignoring the ordering ambiguities
present in the transformation.  The results are shown to correspond to the
correct energy spectrum for the cases that the parent Hamiltonian
constituted a solvable problem. Section~IV begins by examining the
ramifications for the path integral of assuming the existence of new
canonically conjugate variables.  A consistency condition for this change of
variables is derived from the demand for a unit projection operator.  The
quantum canonical transformation is introduced in terms of a generating
function and the forms of the new variables are derived.  The consistency
condition is found to be related to the problem of initial and final
conditions for the new variables, and some of the limitations of the quantum
canonical transformation are revealed.  From the form of the new variables
the explicit form for the Jacobian or anomaly of the transformation is
calculated, and its incorporation into the action of the path integral is
shown.  For a general form of the generating function the anomaly itself is
shown to be a surface term, contributing to the overall prefactor or van
Vleck determinant.  The form of the transformed Hamiltonian is discussed,
and a perturbative proof of the suppression of ordering terms for a cyclic
Hamiltonian is given.  In Sec.~V various systems, some known exactly by
other methods, are evaluated.  These include the velocity-dependent
transformation of the free particle, the linear potential, the transition
from Cartesian to polar coordinates, and the harmonic oscillator.  Finally,
the results are extended to give an approximate solution to the important
case of the time-dependent harmonic oscillator.

\newsection{The Hamiltonian Path Integral}

In this section several aspects of Hamiltonian path integrals that are
relevant to developments later in this paper will be discussed.  While these
aspects may appear at first blush to be unrelated, they will be important
later in this paper to understanding the consequences of canonically
transforming the variables of integration in the path integral.

\subsection{Defining the Path Integral}

The transition element to be analyzed in the remainder of this paper is
given in its one-dimensional form by
\begin{equation}
\label{one} W_{fi} =
\langle \, p_f \, | \exp \left ( - i \hat{H} T / \hbar \right ) |
\, q_i \, \rangle \, .
\end{equation}
The final state, $| \, p_f \, \rangle$, is assumed to be an eigenstate of
the momentum, $\hat{p}$, while the initial state, $| \, q_i \, \rangle$, is
an eigenstate of the position, $\hat{q}$.  The two operators satisfy the
usual algebra $ [ \hat{q} , \hat{p} ] = i \hbar$.  The Hamiltonian
$\hat{H}$ is assumed to be a function of some ordering of $\hat{q}$ and
$\hat{p}$, and its eigenstates, as well as those of $\hat{q}$ and $\hat{p}$,
are determined consistent with any boundary conditions, such as periodicity
in $q$.

$W_{fi}$ is trivial to evaluate if $\hat{H}$ is cyclic, {\it i.e.}, a
function solely of $\hat{p}$.  For such a case it reduces to
\begin{equation}
\label{two}
W_{fi} =
\langle \, p_f \, | \, q_i \, \rangle \,
\exp \left ( -  i H ( p_f ) \, T / \hbar \right )
\, .
\end{equation} The allowed values of the variables $p_f$ and $q_i$ appearing
in the inner product in (\ref{two}) are determined by the boundary
conditions of the original problem, although in one dimension the inner
product for continuous systems takes the general form
\begin{equation}
\label{three}
\langle \, p_f \, | \, q_i \, \rangle =
\frac{1}{\sqrt{2 \pi \hbar}} \exp \left (  - i p_f q_i / \hbar \right ) \, .
\end{equation}
The propagator of the quantum mechanical problem can be derived from result
(\ref{one}). Assuming that the momentum state spectrum is continuous,
the propagator is obtained by a Fourier transform,
\begin{equation}
\label{four}
\langle \, q_f \, | \exp \left ( - i \hat{H} T / \hbar \right )
| \, q_i \, \rangle =
\int \frac{dp_f}{\sqrt{2 \pi \hbar}} \, e^{i p_f q_f / \hbar } \, \langle \,
p_f
\, |
\exp \left ( - i \hat{H} T / \hbar \right ) | \, q_i \, \rangle \; .
\end{equation}
Obviously, a discrete spectrum of momentum eigenstates leads to a Fourier
series.  Once the propagator (\ref{four}) is known, results such as the ground
state energy can be derived.

The Hamiltonian path integral representation of (\ref{one}) may be derived
by using the completeness of the position and momentum eigenstates to perform
a time-slicing argument. This technique is well documented\cite{Swan1}, and
its application here is accomplished by using the unit projection operator
given by
\begin{equation}
\label{five}
\hat{\openone} = \int \frac{dp_j}{\sqrt{2 \pi \hbar}} \, dq_{j+1} \, | \,
q_{j+1} \,
\rangle e^{i q_{j+1} p_j / \hbar} \langle \, p_j \, | \, .
\end{equation}
There is an important subtlety in (\ref{five}).  As it is written it assumes
that the spectra of the states are continuous; however, this will not be the
case in the event that the configuration space of the system is compact or
periodic.  Putting aside such a possibility for the moment, the result of
time-slicing $T$ into $N$ intervals of duration $\epsilon$, where $\epsilon
= T/N$, gives\
\begin{equation}
\label{six}
\langle \, p_j \, | \exp ( - i \epsilon \hat{H} / \hbar ) | \, q_j \,
\rangle =
\frac{1}{\sqrt{2 \pi \hbar}}
\exp \left \{ -  \frac{i}{\hbar} \left [ q_j p_j  + \epsilon
H ( p_j , q_j ) + O ( \epsilon^2 ) \right ] \right \} \, ,
\end{equation}
where the $O( \epsilon^2 )$ terms arise from commutators occurring in the
ordering of the Hamiltonian power series.  This immediately yields the
Hamiltonian path integral recipe for calculating the transition amplitude:
\begin{eqnarray}
\label{seven}
 && W_{fi} = \langle \, p_f \, | e^{- i H T / \hbar } | \, q_i \, \rangle =
\nonumber \\ &&
\frac{1}{\sqrt{2 \pi \hbar}} \int \prod_{j=1}^N
\frac{dp_j}{2 \pi \hbar} \, dq_j \,
\exp \left \{ \frac{i}{\hbar} \sum_{j = 1}^N \left [ - q_j ( p_{j+1} - p_j )
- \rule{0ex}{2.5ex} \epsilon H ( p_j , q_j ) \right ]  - \frac{i}{\hbar}
q_i p_1 \right \}  \, ,
\end{eqnarray}
where $p_{N+1} = p_f$ and the limits $N \rightarrow \infty$ and
$\epsilon \rightarrow 0$ are understood.

\subsection{\protect The Leading Behavior of $\Delta q$}

It is standard practice to assume a continuous form for the path integral by
identifying $\epsilon \rightarrow {\rm dt}$ and $q_j ( p_{j+1} - p_j ) =
{\rm dt} \, q_j \, \dot{p}_j$.  The latter identification is {\em purely
formal}, since $p_{j+1}$ and $p_j$ are independent variables of integration
unrelated by any time evolution.  Even with this formal identification the
action density in the path integral (\ref{seven}) can take the
(semi)standard form, $- q \dot{p} - H$, only if $p_i = 0$ or $q_i = 0$. For
these cases the final term can be written $- i q_i ( p_1 - p_i)$.  This will
be discussed in greater detail in the Sec.~IV.  Another technicality arises
since the argument of the path integral does not satisfy the criteria of a
probability measure unless the time is continued to imaginary values, the
so-called Wick rotation. Otherwise the oscillatory integrands result in
distributions rather than functions.  The Wick rotation will be used and
assumed to yield a sensible measure for all path integrals considered in the
remainder of this paper.

To demonstrate the formal nature of the identification $q_j ( p_{j+1} - p_j
) = {\rm dt} \, q_j \, \dot{p}_j$ as well as derive results that will be
important later in this paper, it will be of use to discuss the leading
behavior in $\epsilon$ of the expectation value of the element $ \Delta q_j
= q_{j+1} - q_j $.  For it to be possible to treat $\Delta q_j$ as $ \dot{q}
\, {\rm dt}$ its expectation value, $ \langle \Delta q_j \rangle_{fi}$, must
be shown to be $O( \epsilon )$.  The behavior of $\langle \Delta q_j \rangle$
is of course a function of the specific form of the Hamiltonian.  However,
if the Hamiltonian is cyclic, then it is always true that $ \langle \Delta
q_j \rangle_{fi} $ is $O ( \epsilon )$. This is easy to demonstrate within
the operator context, where the operator form for Hamilton's equation gives
\begin{equation}
\label{eight}
\Delta \hat{q} (t) = \hat{q} ( t + \epsilon ) - \hat{q} ( t) =
\epsilon \, \frac{i}{ \hbar} [ \hat{H} ( \hat{p} ) , \hat{q} ( t ) ] =
 \epsilon \, \frac{ \partial \hat{ H} ( \hat{p} ) }{ \partial \hat{p} } \, .
\end{equation}
Inserting (\ref{eight}) into (\ref{one}) and using (\ref{two}) immediately
yields
\begin{equation}
\label{nine}
\langle \Delta \hat{q} (t) \rangle_{fi} =
 \epsilon \, \langle \, p_f \, | \, q_i \, \rangle \,
\frac{ \partial  H ( p_f ) }{ \partial p_f }
\exp \left ( -  i H ( p_f ) \, T / \hbar \right )
\, .
\end{equation}

Demonstrating the path integral equivalent of result (\ref{nine}) requires
adding a source term $ K_j \Delta q_j $ to the action. In order to avoid
difficulties with the boundary conditions on $q_j$, the boundary conditions
$K_N = K_0 = 0$ are imposed on the source function.  The expectation
value is then given by
\begin{equation}
\label{ten}
\langle \Delta q_j \rangle_{fi} =
- \left . \frac{i}{\hbar} \frac{ \partial W_{fi} [ K ] }{ \partial K_j }
\right |_{K = 0} \, .
\end{equation}
The next step is to perform the path integral version of integrating by parts
by using the boundary condition on $K$ to rearrange the sum over $j$:
\begin{equation}
\label{11}
\sum_{j=1}^N K_j \Delta q_j =
\sum_{j=1}^N K_j ( q_{j+1} - q_j )
= - \sum_{j=1}^{N} q_j ( K_j - K_{j-1} ) \, \equiv \,
- \sum_{j=1}^N q_j \Delta K_j \, .
\end{equation}
Since $H$ depends only on $p$ all $q$ integrations can now be performed.
Assuming that the range of the $q$ integrals is $\pm \infty$, each
of the $N$ integrations over $q$ yields a Dirac delta,
\begin{equation}
\label{12}
\int dq_j  \, \exp \left \{ - \frac{i}{\hbar} q_j (
p_{j+1} - p_j + \Delta K_j ) \right \}  =
2 \pi \hbar \, \delta ( p_{j+1} - p_j + \Delta K_j )
\, .
\end{equation}
The $p$ variables are now trivial to integrate, giving the result for the
transition element
\begin{equation}
\label{13}
W_{fi} [ K ] = \frac{1}{ \sqrt{ 2 \pi \hbar}}
\exp \left \{ - \frac{i}{\hbar} \left ( p_f q_i + \sum_{j=1}^N \epsilon \, H
( p_f - K_{j-1} ) \right ) \right \}
\, .
\end{equation}
Using (\ref{13}) in (\ref{ten}), along with the result that
\begin{equation}
\label{14}
\lim_{N \rightarrow \infty} \sum_{j=1}^N \epsilon = T \; ,
\end{equation}
reproduces the operator result (\ref{nine}):
\begin{equation}
\label{15}
\lim_{N \rightarrow \infty} \langle \Delta q_j  \rangle_{fi}
=   \frac{\epsilon}{ \sqrt{ 2 \pi \hbar}}  \,
\frac{ \partial  H ( p_f ) }{ \partial p_f }
\exp \left \{ -  \frac{i}{\hbar} [ \, p_f q_i +  H ( p_f ) \, T ] \right \}
\, .
\end{equation}

Result (\ref{15}) does not necessarily follow for non-cyclic Hamiltonians.
The argument used to derive (\ref{15}) can be applied to the harmonic
oscillator action to show that $\langle ( \Delta q_j )^2
\rangle_{fi}$ is $O ( \epsilon )$. This is easily seen from the Gaussian
nature of the Wick-rotated integrations. The $q$ integration results in
\begin{equation}
\int dq_j \, \exp \left [ - \epsilon \frac{1}{2} {q_j}^2 + q_j ( \Delta K_j
+ \Delta p_j ) \right ]
= \sqrt{ \frac{2 \pi}{ \epsilon } } \exp \left [ - \frac { ( \Delta p_j
+ \Delta K_j )^2}{ \epsilon} \right ] \, .
\end{equation}
The $\Delta K_j $ dependence may be removed from this term by translating
the $p_j$ variables according to $p_j \rightarrow p_j + K_j$.  Doing so
changes the ${p_j}^2$ term in the exponential of the path integral according
to
\begin{equation}
- \epsilon \frac{1}{2} {p_j}^2 \rightarrow
- \epsilon \frac{1}{2} {p_j}^2 - \epsilon p_j K_j
- \epsilon \frac{1}{2} {K_j}^2
\end{equation}
It is then clear that the second derivative of the resulting function with
respect to $K_j$ will result in a term $O ( \epsilon )$.  In an identical
manner it is possible to show that the expectation value of $\Delta p_j =
p_{j+1} - p_j $ vanishes if the Hamiltonian is cyclic.  This is the quantum
mechanical equivalent of the classical Hamilton's equation
\begin{equation}
\dot{p} = - \frac{ \partial H}{ \partial q} \, .
\end{equation}

As a result of (\ref{15}) it is possible to use a perturbative argument to
show that certain types of terms in the Hamiltonian of the path integral are
suppressed in the limit $N \rightarrow \infty$.  The Hamiltonian under
consideration has the form
\begin{equation}
\label{16}
H = H_{cl} ( p , q ) + H_\Delta ( \Delta q , q ) \, ,
\end{equation}
where all terms in $H_\Delta$ have at least one positive power of $\Delta q$
and $H_{cl}$ is the Hamiltonian inherited from the classical system.  Such a
Hamiltonian has no classical counterpart, since terms of the form $H_\Delta$
would be suppressed \cite{Prok}.  If $H_{cl}$ is cyclic it can be shown that
the terms $H_\Delta$ do not contribute to the path integral in the limit $N
\rightarrow \infty$.  The argument is similar to the one used to demonstrate
(\ref{15}). The contribution of the terms $H_\Delta$ is written as a
perturbation series using $H_{cl}$ as the basis Hamiltonian.  This is
accomplished by adding the source terms $ \epsilon K_j
\Delta q_j$ and $ \epsilon J_j q_j$ to the action without $H_\Delta$ to give
the function $W_{fi} [ K , J ]$.  The perturbation series representation of
the original transition element is then defined as
\begin{equation}
\label{18}
\left . \exp \left \{ - \frac{i}{\hbar} \sum_{j=1}^{N} \epsilon \, H_\Delta
\left ( \frac{ \hbar }{ i \epsilon} \frac{ \partial}{ \partial K_j} \, ,
\frac{ \hbar}{ i \epsilon} \frac{ \partial}{  \partial J_j} \right ) \right \}
W_{fi} [ K , J ] \right |_{K,J = 0} \, .
\end{equation}
The function $W_{fi} [ K , J ]$ is readily evaluated to give
\begin{equation}
\label{19}
W_{fi} [ K ] = \frac{1}{ \sqrt{ 2 \pi \hbar }}
\exp \left \{ - \frac{i}{\hbar} \left ( p_f q_i + \sum_{j=1}^N \epsilon \, H
( p_f - \epsilon K_{j-1} - \sum_{l = 1}^{j} \epsilon \, J_l ) \right ) \right
\}
\, .
\end{equation}
While the $ \sum \epsilon J $ term results in an integral in the limit
$\epsilon \rightarrow 0$, it is clear that the term $ \epsilon K$ is
suppressed relative to the other terms by a factor of $T/N$.  The
derivatives with respect to $K$ are also suppressed by this factor as well,
showing that terms of the form $H_\Delta$ do not contribute to the
perturbation series. For the case that the basis Hamiltonian is cyclic such
terms can therefore be discarded.  In effect, this perturbative argument
substantiates the general intuition that, for a cyclic Hamiltonian, $\Delta q$
can be replaced by
$\epsilon \dot{q}$, where $\dot{q}$ is finite.  Any resulting terms with
factors of $O( \epsilon^2 )$ or greater can then be suppressed.

Since perturbative arguments are fraught with pitfalls and loopholes, it is
worth checking this result for exactly integrable cases.  For example, the
path integral whose Hamiltonian is given by
\begin{equation}
H = \frac{1}{2} p^2 + \lambda q \Delta q
\end{equation}
can be shown to reduce to the standard cyclic result (\ref{two}) with all
terms proportional to $\lambda$ suppressed by an additional factor of
$\epsilon^2$.  Terms of the form $ q \Delta p$ or $p \Delta p$ can be
integrated exactly to find the standard cyclic result in the limit $\epsilon
\rightarrow 0$.  However, there is at least one important set of cases not
covered by this perturbative argument involving terms with quadratic powers
of $p$.  For example, if the term $ f(q) \Delta q \, p^2$ occurs in the
Hamiltonian, its contribution cannot be discarded.  It is not difficult to
see the mechanism for this by examining the Hamiltonian
\begin{equation}
\label{20a}
H = \frac{1}{2} p^2 + \frac{1}{2} \Delta q \, f(q) p^2 =
\frac{1}{2} ( 1 + f(q) \Delta q) p^2 \, .
\end{equation}
The $p$ integrations in the Wick-rotated path integral are Gaussian, and
take the general form
\begin{eqnarray}
\label{20}
&&\int \frac{dp_j}{2 \pi \hbar }
\exp \left \{ - \frac{\epsilon}{\hbar} \left [ \frac{1}{2} ( 1 + f(q_j)
\Delta q_j ) {p_j}^2 - p_j \frac{\Delta q_j}{\epsilon} \right ] \right
\} \nonumber \\
&& =
\sqrt{ \frac{ 1 }{ 2 \pi \hbar \epsilon ( 1 + f(q_j) \Delta q_j ) } } \;
\exp \left [ \frac{ ( \Delta q_j )^2}{ 2 \hbar \epsilon ( 1 + f(q_j) \Delta
q_j)}
\right ]
\, .
\end{eqnarray}
If $\Delta q$ remains proportional to some positive power or root of
$\epsilon$, then the $\Delta q$ term in the denominator of the exponential
can be discarded due to the factor of $\epsilon$ present in the
denominator.  However, the terms in the prefactor may contribute to the path
integral.  This follows from the fact that the prefactor terms can be written
\begin{equation}
\label{21}
\frac{1}{ \sqrt{ 1 + f(q_j) \Delta q_j } }
= \exp \left [ - \frac{1}{2} \ln ( 1 + f(q_j) \Delta q_j ) \right ]
\approx \exp \left [ - \frac{1}{2} f(q_j) \Delta q_j \right ]
\, .
\end{equation}
Even if $\Delta q_j \approx \epsilon$, the infinite sum in which (\ref{21})
becomes embedded can result in a nontrivial contribution since $N \epsilon
\rightarrow T$.  The upshot of result (\ref{21}) is to transmute the
original interaction term $ f(q) \Delta q \, p^2$ into an effective
velocity-dependent potential in the path integral when all momenta have been
integrated. This velocity-dependent potential appears proportional to
$\hbar$, since (\ref{20}) can be written
\begin{equation}
\label{22}
\sqrt{ \frac{ 1 }{ 2 \pi \hbar \epsilon ( 1 + f(q_j) \Delta q_j ) } } \;
\exp \left [ \frac{ ( \Delta q_j )^2}{ 2 \hbar \epsilon ( 1 + f(q_j)
\Delta q_j)} \right ] \, \approx \, \sqrt{ \frac{ 1 }{ 2 \pi \hbar  \epsilon} }
\; \exp \left \{ \frac{ \epsilon }{ \hbar } \left [
\frac{1}{2} {\dot{q}}^2 + \frac{1}{2} \hbar f(q) \dot{q} \right ] \right \} \,
,
\end{equation} where the standard path integral notation $\Delta q =
\epsilon \dot{q} $ has been used.  Result (\ref{22}) is consistent with the
idea that the classical Hamiltonian (\ref{20a}) would receive no
contribution from such a potential.  If it is to give a nontrivial
contribution to the quantum mechanical theory, it must be equivalent to a
term in the action of $O( \hbar )$ or higher.  Clearly, similar results can
be obtained for other Gaussian-like terms for specific choices of the cyclic
Hamiltonian.  A discussion of possible terms that may contribute is given by
Prokhorov\cite{Prok}.

It is worth noting for later reference that, if the Hamiltonian is cyclic,
the path integral (\ref{seven}) can be evaluated exactly by translating the
variables of integration by the classical solutions to Hamilton's equations
consistent with the boundary conditions $q(t = 0) = q_i$ and $p (t = T) =
p_f$, given by
\begin{equation}
\label{23}
p_c ( t ) = p_f \, , \; \; q_c (t) = q_i + \frac{ \partial H ( p_f )}{ \partial
p_f } t \; .
\end{equation}
This is possible because the difference between adjacent time-slices {\em
does} reduce to the derivative for a classical function, {\it i.e.}, $ q_c (
t_{j+1} ) = q_c ( t_j + \epsilon ) \rightarrow q_c (t_j) + \epsilon
\dot{q}_c (t_j)$.  Performing an integration by parts similar to (\ref{11})
reduces the translated path integral (\ref{seven}) to
\begin{eqnarray}
&& \label{24} \frac{1}{ \sqrt{ 2 \pi \hbar }}
\exp \left \{ -  \frac{i}{\hbar} [ \, p_f q_i +  H ( p_f ) \, T ] \right \}
\times \nonumber \\
&& \int \prod_{j=1}^N
\frac{dp_j}{2 \pi \hbar} \, dq_j \,
\exp \left \{ \frac{i}{\hbar} \sum_{j = 1}^N \left [ - q_j ( p_{j+1} - p_j )
- \rule{0ex}{2.5ex} \epsilon
\frac{1}{2} {p_j}^2 \frac{ \partial^2 H ( p_c (t_j) )}{ \partial {p_c (t_j)}^2
} - \ldots \right ]  \right \}  \, ,
\end{eqnarray}
where the ellipsis refers to higher order terms present in the expansion of
$H ( p_c (t_j) + p_j ) $ around $p_j$ and, because of the translation of
variables, $p_{N+1} = 0$.  It is precisely the latter result that reduces
the translated path integral appearing in (\ref{24}) to unity when all
integrations are performed, a fact that is apparent when result (\ref{12})
is examined for the case $K = 0$.  Therefore, the only surviving factor in
(\ref{24}) is the exponential of the classical action evaluated along the
classical trajectory (\ref{23}).

\subsection{Discrete Spectrum Path Integrals}

Another relevant point regards the case where the allowed values of the
momentum or energy constitute a denumerably infinite set rather than a
continuous variable.  Such a result is common in quantum mechanical systems,
occurring in bound state spectra and in systems where the configuration
space is compact or periodic boundary conditions are enforced.  In wave
mechanics the discrete spectrum can arise from demanding either that the
bound state wave-function is normalizable or that the wave-function or its
derivative vanishes on some boundaries.  It is natural to expect that the
measure of the path integral for such a system would differ from its ``free''
counterpart (\ref{seven}).  However, it is often the case that the path
integral representation of the transition amplitude (\ref{one}) for such a
system is identical to the continuous result (\ref{seven}).  This outcome is
well-known within the context of specific systems \cite{Schul3}. Since
this aspect of path integrals is relevant to canonical transformations, the
general derivation of the range of integrations will be sketched for the
specific case of a {\em free} particle constrained to be in a
one-dimensional infinite square well.

The position eigenstates range from $-a$ to $a$, while the momentum
eigenstates, $ | \, n \, \rangle$, are discrete and indexed by an integer
$n$.  Unit projection operators are given by
\begin{equation}
\label{25}
\hat{\openone} = \int_{-a}^a dq \, | \, q \, \rangle \langle \, q \, | \, ,
\; \;
\hat{\openone} = \sum_{n = -\infty}^\infty | \, n \, \rangle \langle \, n \, |
\, ,
\end{equation}
while the inner product is given by
\begin{equation}
\label{26}
\langle \, q \, | \, n \, \rangle = \frac{1}{ \sqrt{ 2 a } }
\exp \left ( \frac{ i \pi n q}{a} \right ) \, .
\end{equation}
Of course, the physical energy eigenstates are linear combinations of $| \,
n \, \rangle$ and $| \, -n \, \rangle$ consistent with the boundary
conditions.  The time-slicing argument that was used to construct
(\ref{seven}) can be revisited using (\ref{25}) and (\ref{26}) to obtain
\begin{eqnarray}
\label{27}
W_{fi} & = & \langle \, n_f \, | e^{- i \hat{H} T / \hbar}
| \, q_i \, \rangle \nonumber \\
& = & (2a)^{- (N+1)/2 } \sum_{n_1, \ldots, n_N} \int_{-a}^{a}
d q_1 \cdots d q_N \, \times \nonumber \\
&& \exp \left \{ - \frac{i}{\hbar} \sum_{j=1}^{N} \left [
\frac{ n_j \pi \hbar }{ a } ( q_j - q_{j-1} )
- \epsilon H ( n_j ) \right ]  - \frac{ i n_f \pi q_N}{a} \right \} \, ,
\end{eqnarray}
where $q_0 = q_i$.

Result (\ref{27}) can be rewritten using the Poisson resummation technique,
which begins by using the identity
\begin{equation}
\label{28}
\sum_{n = -\infty}^{\infty} f(n) = \sum_{k = -\infty}^\infty
\int_{- \infty}^\infty dn \, f(n) \, e^{i 2 \pi k n} \, .
\end{equation}
Using (\ref{28}) and making the obvious definition $p_j = n_j \pi \hbar / a$
allows (\ref{27}) to be written as
\begin{eqnarray}
\label{29}
W_{fi} & = & \frac{1}{ \sqrt{2a} }
\int_{-\infty}^\infty \frac{dp_1}{2 \pi \hbar} \cdots \frac{ dp_N }{ 2 \pi
\hbar} \int_{-a}^a dq_1 \cdots dq_N \sum_{k_1,\ldots,k_N} \times \nonumber \\
& & \exp \left \{ - \frac{i}{\hbar} \sum_{j=1}^{N} \left [
p_j ( q_j + 2a k_j - q_{j-1} )
- \epsilon H ( p_j ) \right ]  - \frac{i}{\hbar} p_f q_N \right \} \, .
\end{eqnarray}
Because the Hamiltonian is independent of $q$ and the sums over the $k_j$
are infinite, the sums may be absorbed by extending the range of the $q_j$
integrations.  However, this is contingent on the fact that
\begin{equation}
\label{30}
\exp \left \{ - \frac{i}{\hbar} p_f ( q_N + k_N 2a) \right \}
= \exp \left \{ - \frac{i}{\hbar} p_f q_N \right \} \, ,
\end{equation}
which holds as long as $p_f = n_f \pi \hbar / a$ and $n_f$ is an integer.
Because the wave-mechanical solution to the problem was used to derive the
path integral form, it is clear that condition (\ref{30}) holds.  The final
form of the path integral is given by
\begin{equation}
\label{31}
W_{fi} =
\frac{1}{ \sqrt{2a} }
\int_{-\infty}^\infty \frac{dp_1}{2 \pi \hbar} \cdots \frac{ dp_N }{ 2 \pi
\hbar} \, dq_1 \cdots dq_N
\exp \left \{ - \frac{i}{\hbar} \sum_{j=1}^{N} \left [
q_j ( p_{j+1} - p_j )
- \epsilon H ( p_j ) \right ]  - \frac{i}{\hbar} p_1 q_i \right \} \, ,
\end{equation}
where the limits on the $q_j$ integrations is now $\pm \infty$.  Apart from
the overall factor of $ ( 2 a )^{- 1/2 }$, result (\ref{31}), with its
ambiguous symbol $p_{N+1} = p_f$, is formally identical in its measure and
action to the free case (\ref{seven}). In that sense information about the
system has been lost in the transition from (\ref{27}) to (\ref{31}) since
the form (\ref{31}) does not specify a discrete spectrum for $p_f$.  {\it A
priori} knowledge of the momentum spectrum is required in order that the
discrete form of the Fourier transform, rather than the continuous form, is
employed to obtain the propagator (\ref{four}).

\subsection{Fourier Methods for Evaluating Path Integrals}

A final aspect of importance regarding the Hamiltonian path integral is the
method that does allow the action in the path integral to be manipulated as
if the formal time derivative was a true derivative.  The $q_j$ and $p_j$
variables are first translated by a classical solution to Hamilton's
equations of motion consistent with the boundary conditions.  This means
using classical solutions for both $p$ and $q$ that satisfy the conditions
\begin{equation}
p_c ( t = T ) = p_f \, , \; \; q_c ( t = 0) = q_i \, .
\end{equation}
Because there is no initial condition for $p$ or final condition for $q$ in
the original form of the transition element, the translated $p$ and $q$
variables do not necessarily vanish at {\em both} $t = 0$ and $t = T$.
Consistent with these boundary conditions, the $2N$ fluctuation variables
$q_j$ and $p_j$ are written as Fourier expansions in terms of $2N$ new
variables $q_n$ and $p_n$,
\begin{eqnarray}
\label{33}
q_j & = & \sum_{n = 0}^{N-1} q_n \sin \left ( \frac{ (2n +1) \pi t_j }{ 2T }
\right
)   \\
\label{33.2}
p_j & = & \sum_{n=0}^{N-1} p_n \left ( \frac{(2 n +1) \pi}{2 T } \right )
\cos \left ( \frac{ (2n+1) \pi t_j }{ 2T } \right
) \, .
\end{eqnarray}
The expansions of (\ref{33}) satisfy the proper translated quantum boundary
conditions $p ( t = T) = 0$ and $q ( t = 0) = 0$, but are arbitrary at the
remaining endpoints in order to accommodate the quantum nature of the
coordinates.  This is an outgrowth of the uncertainty principle for canonical
variables, since the uncertainty principle forces $q_f$ to be undefined if
$p_f$ is exactly known, with a similar relation between $q_i$ and $p_i$.
However, the formal derivatives in the path integral now become true
derivatives, since to $O (\epsilon)$
\begin{equation}
\label{34}
q_{j+1} - q_j \rightarrow \epsilon \sum_{n = 1}^N q_n \frac{( 2n +1) \pi}{2 T}
\cos \left ( \frac{ (2n+1) \pi t_j }{2 T } \right ) \, .
\end{equation}
The measure is rewritten in terms of integrations over the coefficients of
the Fourier expansions. This change of variables is accompanied by a
Jacobian that is nontrivial, but one that can be inferred by forcing the new
path integral to yield the same results as the configuration space measure
version discussed in the previous part of this section.  In addition to the
usual Wick rotation $T \rightarrow - i T$, the Hamiltonian path integral
also requires $p_n \rightarrow - i p_n$.  The case of an arbitrary cyclic
Hamiltonian is particularly easy since the integrations over the $q_n$
variables yield a factor of the form
\begin{equation}
\label{35}
\left ( \frac{ 8 T}{ \pi^2 } \right )^N \prod_{n=0}^{N-1} ( 2n+1)^{-2}
\delta( p_n ) \, ,
\end{equation}
from which the Jacobian is inferred to be
\begin{equation}
\label{36}
J = \left ( \frac{  \pi^2 }{ 8 T } \right )^N [ ( 2N - 1)!! ]^2 \, .
\end{equation}

The validity of this procedure can be tested on the harmonic oscillator
transition element.  There the classical solutions consistent with the
boundary conditions are
\begin{equation}
\label{37}
q_c ( t ) = A \sin ( \omega t + \delta ) \, , \; \;
p_c ( t ) = m \omega A \cos ( \omega t + \delta ) \, ,
\end{equation}
where
\begin{equation}
\label{38}
A = q_i \, {\rm csc \,} \delta \, , \; \;
\cot \delta = \frac{ p_f {\rm \, sec \,} (\omega T) }{ m \omega q_i }
- \tan ( \omega T ) \, .
\end{equation}
Using (\ref{37}) and (\ref{38}) in the harmonic oscillator action yields the
result
\begin{equation}
\label{39}
\int_0^T {\rm dt} \, {\cal L} =  q_i p_f {\rm \, sec \, } ( \omega T )
- \frac{1}{2} m \omega {q_i}^2 \tan ( \omega T )
- \frac{1}{2} \frac{ {p_f}^2 }{ m \omega } \tan ( \omega T ) \, .
\end{equation}
The remaining translated action reduces to Gaussians in both $p_n$ and
$q_n$.  Performing the integrations, combining the result with the Jacobian
(\ref{36}), and undoing the Wick rotation yields the prefactor
\begin{equation}
\label{40}
\lim_{N \rightarrow \infty} \prod_{n=0}^{N-1}
\left ( 1 - \frac{ 4 \omega^2 T^2 }{ (2 n + 1)^2 } \right )^{-\frac{1}{2}}
= \frac{ 1 }{ \sqrt{ \cos \omega T } } \; .
\end{equation}
Combining results (\ref{39}) and (\ref{40}) yields the correct form for the
transition element (\ref{one}) for the harmonic oscillator.  In Sec.~IV the
ramifications of the quantum nature of the coordinate fluctuations for the
boundary conditions of the canonically transformed coordinates will be
discussed, and the results of this subsection will be used to define
restrictions on the validity of the canonically transformed path integral.

\newsection{Canonical Transformations}

A classical canonical transformation is one from the coordinates $(q,p)$ to
a new set of coordinates $(Q,P)$ such that the Poisson bracket structure, or
equivalently the volume of phase space, is preserved.  For convenience and
consistency only canonical transformations of the third kind \cite{Gold}
will be considered in the remainder of this paper, and these are defined by
a choice for the generating function of the general form $F( p , Q , t)$.
At the classical level the new variables are determined by solving the
system of equations given by
\begin{equation}
\label{45}
q = - \frac{ \partial F ( p , Q , t)}{ \partial p} \, ,
\; \; P = - \frac{ \partial F (p, Q, t)}{ \partial Q} \, .
\end{equation}
It is important to remember that $Q$ and $p$ are treated as independent in
the definitions of the new coordinates given by (\ref{45}).  However, the
proof that the Poisson bracket structure is preserved depends on the
identities obtained by differentiating (\ref{45}) and using the fact that $Q
= Q(q,p)$. For example, it follows that
\begin{equation}
\label{45.1}
1 = \frac{ \partial q}{ \partial q} = - \frac{ \partial^2 F(p,Q,t)}
{\partial Q \, \partial p} \frac{ \partial Q ( q , p , t ) }{ \partial q}
\, .
\end{equation}
It is assumed that the equations of (\ref{45}) are well-defined and can be
solved to yield $Q(q,p,t)$ and $P(q,p,t)$, or inverted to obtain $q(Q,P,t)$
and $p(Q,P,t)$.  The action is transformed according to
\begin{eqnarray}
\label{46}
& & \int_0^T dt \, \left [ - q \dot{p} - H ( p, q) \right ] \nonumber \\
& & = \int_0^T dt \, \left [ P \dot{Q} - \tilde{H} ( P , Q ) + \frac{dF}{dt}
\right ]  \nonumber \\
&  & = F ( p_f , Q_f, t_f ) - F( p_i , Q_i, t_i ) +
\int_0^T dt \, \left [ P \dot{Q} - \tilde{H} ( P , Q ) \right ] \, ,
\end{eqnarray}
where
\begin{equation}
\label{47}
\tilde{H} ( P , Q )  = H ( p(Q,P,t) , q(Q,P,t) ) +
\frac{ \partial F(p(P,Q),Q,t)}{\partial t} \; .
\end{equation}

At the classical level there is no difficulty in obtaining initial and final
values for both variables $q$ and $p$ since it is assumed that Hamilton's
equations can be solved to obtain classical solutions consistent with any
possible pair of boundary conditions over the arbitrary time interval $T$.
The two unspecified endpoint values of the variables are simply those given
by the classical solutions at the respective endpoint times.  However, if
canonical transformations are to be employed in a path integral setting in a
manner similar to the classical result, it is necessary to deal with the
quantum mechanical version of this problem, and there is no {\it a priori}
reason to expect that the classical definition is consistent with the
quantum mechanical transition amplitude (\ref{one}).  This will be discussed
in detail in Sec.~IV.

It is apparent at the classical level that the values of the generating
function evaluated at the endpoints, {\it i.e.}, the surface terms,
correspond to a piece of the minimized original classical action not
determined by the minimized transformed action.  This is demonstrated by
examining the well-known canonical transformation to cyclic coordinates for
the harmonic oscillator.  Using the generating function
\begin{equation}
\label{48}
F( p , Q ) = - \frac{p^2}{2 m \omega} \tan Q
\end{equation}
gives
\begin{equation}
\label{49}
Q = \arctan \frac{ m \omega q}{ p } \, , \; \; \omega P =
\frac{p^2}{2 m} + \frac{1}{2} m \omega^2 q^2 \, ,
\end{equation}
so that transformed Hamiltonian is $\tilde{H} = \omega P$.  It follows that
the transformed action vanishes when evaluated along the classical
trajectory $Q_c = \omega t + Q_i$ and $P_c = P_f = P_i$.  Therefore, the
value of the original action along the classical trajectories $q_c$ and
$p_c$ is contained entirely in the endpoint contributions of the generating
function.  An explicit calculation using the classical solutions (\ref{37})
in (\ref{48}) with $Q_i = \delta$ and $Q_f = \omega T + \delta$ verifies
that the generating function endpoint values reproduce result (\ref{39}).
Because the value of the action along the classical trajectory is the phase
of the quantum transition amplitude (\ref{one}) in the WKB approximation,
the value of knowing the form of the generating function for a canonical
transformation to cyclic coordinates becomes apparent.  Such a classical
generating function already gives considerable information regarding the
quantum transition amplitude.

However, it is important to note that the choice of $\omega$ appearing in
the transformed Hamiltonian $\tilde{H}$ is arbitrary.  Using the generating
function
\begin{equation}
\label{50}
F( p , Q ) =
- \frac{p^2}{2 m \omega} \tan \left ( \frac{\omega}{\omega^\prime}
Q \right )
\end{equation}
transforms the Hamiltonian to $\omega^\prime P$.  While the solution for $Q$
becomes $ Q = \omega^\prime t + Q_i$, this has no effect on the solution for
the original variable $q$ because of the offsetting factors in the
generating function.  This is merely a reflection of the fact that scaling
$Q$ can be offset by scaling $P$ and $m$ or $\omega$ when the Hamiltonian is
cyclic.  The generating function can also undergo arbitrary translations of
the $Q$ variable as well, which are offset simply by choosing a different
value for $Q_i$ in the classical solution.

The classical harmonic oscillator solution can be generalized to power
potentials of the form
\begin{equation}
\label{51}
H = \frac{p^2}{2m} + \frac{1}{n} m \lambda^n q^n \; ,
\end{equation}
where $\lambda$ is a constant with the natural units of inverse length.
Hamiltonians of the form (\ref{51}) are rendered cyclic by using the
generating function
\begin{equation}
\label{52}
F = - \frac{1}{2 \alpha}
\left ( \frac{p^2}{m \lambda} \right )^{\alpha} f^{\gamma} (Q)\, ,
\end{equation}
where $\alpha = (n+ 2)/2n$ and $\gamma = - 2/n$.  Using this generating
function gives
\begin{eqnarray}
\label{53}
q & = & \frac{p^{2 \alpha -1}}{ (m \lambda)^\alpha}  f^{\gamma} (Q) \; , \\
\label{54}
P & = & \frac{\gamma}{2 \alpha} \frac{ p^{2 \alpha} }{(m \lambda)^\alpha}
f^{(\gamma - 1)} (Q) \frac{\partial f(Q)}{\partial Q} \, .
\end{eqnarray}
Substituting (\ref{53}) into the original Hamiltonian gives
\begin{equation}
\label{55}
H = p^2 \left [ \frac{1}{2m} + \frac{1}{n \lambda} \left (
\frac{ \lambda }{ m} \right )^{\frac{n}{2}} f^{\gamma n} (Q) \right ]
\, .
\end{equation}
Using (\ref{54}) shows that (\ref{55}) reduces to
\begin{equation}
\label{56}
\tilde{H} = \omega P^\beta \, ,
\end{equation}
where $ \beta = 2n / (n+2) = 1 / \alpha$, if $f(Q)$ is chosen to satisfy the
first-order differential equation
\begin{equation}
\label{57}
\frac{ \partial f(Q) }{ \partial Q} =
\frac{ 2 \alpha}{ \gamma } \left ( \frac{ \lambda}{ 2 \omega}
\right )^\alpha
\left [ f^2 (Q) + \frac{ 2m}{ n \lambda }
\left ( \frac{ \lambda }{ m } \right )^{\frac{n}{2}}
\right ]^\alpha \, .
\end{equation}
While a particular value for the $\omega$ in (\ref{56}) can be chosen, such
a choice is arbitrary in the same way that the $\omega^\prime$ in (\ref{50})
is arbitrary.  This arbitrariness in scale is similar to that which appears
in equivariant cohomology approaches to the same problem \cite{Dykstra}.
The sign for $\omega$ is determined from the range of the original
Hamiltonian, and can be either positive or negative if the original
Hamiltonian was such that $n <0$.  Choosing a negative sign for $\omega$
will affect the final form of expression (\ref{57}).  However, it is
important to note that if $n$ is odd, the range of the original Hamiltonian
is $- \infty$ to $\infty$.  This will introduce difficulties in maintaining
the range of the Hamiltonian in some cases.  This will be demonstrated for
the specific case of a linear potential in Sec.~V.B.

Equation (\ref{57}) can be formally solved by integration, so that
\begin{equation}
\label{58}
\frac{ 2 \alpha}{ \gamma } \left ( \frac{ \lambda}{ 2 \omega}
\right )^\alpha
( Q - Q_i ) = \int \frac{ df }{ \left [ f^2  + \zeta^2
\right ]^\alpha} \, ,
\end{equation}
where
\begin{equation}
\label{59}
\zeta^2 =
\frac{ 2m}{ n \lambda }
\left ( \frac{ \lambda }{ m } \right )^{\frac{n}{2}} \, .
\end{equation}
The right hand side of (\ref{58}) is, up to the factor on the left hand side,
the functional inverse of $f$, written $g$, so that $g ( f ( Q) ) = Q$.
Therefore, inverting (\ref{58}), where possible, yields the function $f(Q)$
appearing in the canonical transformation.  However, even if the expression
generated by (\ref{58}) cannot be exactly inverted, it can still be used to
determined the classical form for $Q(q,p)$ in the following way.  Form
(\ref{53}) shows that
\begin{equation}
\label{60}
Q = g \left (
\left [
\frac{ m^\alpha \lambda^\alpha q }{ p^{2 \alpha - 1} }
\right ]^{\frac{1}{\gamma}}
\right ) \, ,
\end{equation}
so that the result of the integral (\ref{58}), written as a function of
$f$, must coincide with result (\ref{60}).  Therefore, substituting
\begin{equation}
\label{61}
f = \left ( \frac{ m^\alpha \lambda^\alpha q }{ p^{2 \alpha - 1} }
\right )^{\frac{1}{\gamma}}
\end{equation}
into the result of the integration in (\ref{58}) gives $Q = Q( q, p)$.  It
is easy to show that the choices $n = 2$, $\omega = \lambda$, and $Q_i =
\pi / 2$ reproduce the harmonic oscillator generating function (\ref{50}).
However, the cyclic form (\ref{56}) for the transformed Hamiltonian is not
unique since a second transformation using the generating function $F = -
f(P) Q^\prime $ results in a Hamiltonian that is an arbitrary function of
$P^\prime = f(P)$ alone.  Nevertheless, in any cyclic Hamiltonian the
remaining variable is some function of the original Hamiltonian, {\it i.e.},
$P = P ( H ( p, q) )$.

Any attempts to use these results within the quantum mechanical context are
immediately beset by ordering problems.  While the classical Poisson bracket
of $Q$ and $P$ remains unity, the original algebra of $q$ and $p$, coupled
with the transcendental nature of the transformation, results in the
commutator of $Q$ and $P$ being poorly defined.  To lowest order in $\hbar$
it is true that $ [ Q , P ] = i \hbar$, but additional powers appear that
are dependent on the ordering convention chosen for the expansion of the
transcendental functions.  In order to preserve the commutation relations it
is necessary to institute a unitary transformation of the original operator
variables, rather than a canonical transformation.  Anderson \cite{Anderson}
has discussed enlarging the Hilbert space of the original theory to
accommodate non-unitary transformations that alter the commutation
relations.  Although some of the results obtained in such an approach are
similar to those of canonical transformations, this is a fundamentally
different approach to solving the equations of motion.  As a result, it will
not be discussed here.

These ordering ambiguities can be demonstrated by examining the canonical
transformation from Cartesian to spherical coordinates in two dimensions.
The generating function for this transformation is given by
\begin{equation}
\label{SC1}
F = - p_x r \cos \theta - p_y r \sin \theta \, ,
\end{equation}
and this yields the standard result $x = r \cos \theta$, $y = r \sin \theta$,
$ P_r = p_x \cos \theta + p_y \sin \theta$, and $P_\theta = - p_x r \sin
\theta + p_y r \cos \theta$.  In order to invert these equations it is
necessary to choose an ordering convention for the operators.  The most
reliable of these is Weyl ordering, which symmetrizes all non-commuting
operators.  The result is
\begin{eqnarray}
p_x & = & \cos \theta \, P_r - \frac{ \sin \theta}{2r} P_\theta
- P_\theta \frac{ \sin \theta}{2r}
\\
p_y & = & \sin \theta \, P_r + \frac{\cos \theta}{2r} P_\theta
+ P_\theta \frac{\cos \theta}{2r} \, .
\end{eqnarray}
Using the commutators for the spherical coordinates yields
\begin{equation}
\label{SC2}
{p_x}^2 + {p_y}^2 = {P_r}^2 + \frac{1}{r^2} {P_\theta}^2 + \frac{i \hbar}{r}
P_r - \frac{\hbar^2}{ 4 r^2} \, ,
\end{equation}
showing that this transformation takes a cyclic Hamiltonian into a non-cyclic
Hamiltonian.  It is not difficult to see that the $O(\hbar)$ term is
essential to maintaining self-adjointness of the Hamiltonian when written in
terms of spherical coordinates.  This can be seen from an integration by
parts for the expectation value of the Hamiltonian in spherical coordinates
\cite{Swan2}.  Of course, such a term is not generated by the classical
transformation, giving further evidence that classical canonical
transformations and their quantum counterparts may differ by terms that are
functions of $\hbar$.  The numerical factor appearing before the $O (
\hbar^2 )$ term in (\ref{SC2}) is a function of the Weyl ordering chosen for
the original operator expressions.

Nevertheless, because $[ Q, P] \approx i \hbar$, it is interesting to treat
the new variables as if they were canonically conjugate quantum variables
and pursue the solution of the transformed system (\ref{56}).
Alternatively, one could start with the Hamiltonian (\ref{56}), enforce the
exact commutator $[ Q, P] = i \hbar$, and solve for the energy levels of the
system. While it is clear that the previously mentioned ordering problems
prevent this solution from being that of the original system that led to the
cyclic Hamiltonian, such a solution can serve as an approximation to
$O(\hbar^2)$ of the original Hamiltonian. This solution can be found in a
formal manner by assuming a discrete spectrum, {\it i.e.}, bounded from
below, and defining the creation and annihilation operators
\begin{equation}
\label{62}
a^\dagger = e^{i Q \hbar^{\alpha -1} }
\sqrt{ (P - \delta) \hbar^{- \alpha} }  \, , \; \;
a = \sqrt{ (P - \delta)\hbar^{-\alpha }} \, e^{- i Q \hbar^{\alpha -1 } } \, .
\end{equation}
Using the commutator $ [ Q , P ] = i \hbar $ gives $[ a , a^\dagger] =
1$ regardless of the value of $\delta$.  Since $\hbar^\alpha a^\dagger a = P
- ( 1 + \delta) \hbar^\alpha$, the Hamiltonian (\ref{56}) becomes $H = \hbar
\omega ( a^\dagger a + 1 + \delta )^\beta$.  Defining a ground state $ | \,
0 \, \rangle$ by the relation $ a | \, 0 \, \rangle = 0$, it follows that
the excitations of the system are obtained by applying suitably normalized
factors of $a^\dagger$ to the ground state, leading to an energy spectrum $
E_n = \hbar \omega ( n + 1 + \delta)^\beta$.

The arbitrariness of $\delta$ can be used to offset the ordering ambiguities
in the canonical transformation generated by the original algebra of $q$ and
$p$.  The simple harmonic oscillator solution demonstrates this aspect.
Using (\ref{49}) and ignoring commutators of $q$ and $p$ shows that the
annihilation operator of (\ref{62}) contains the factor
\begin{equation}
\label{63}
e^{- i Q} =  - i \sqrt{ \frac{ m \omega }{ 2 P } } \left (
q + \frac{ i p}{ m \omega} \right ) \, .
\end{equation}
The term in the parentheses in (\ref{63}) is, up to a factor, the standard
annihilation operator associated with the harmonic oscillator.  It is also
true that ignoring the ordering ambiguities has resulted in an expression
that does not satisfy $e^{-iQ} e^{iQ} = 1$ at the quantum level, exposing the
formal nature of the manipulations that led to (\ref{63}).  Choosing $\delta
= - 1/2$ reproduces the correct harmonic oscillator energy spectrum.

It is important to determine if the general form for the bound state
spectrum of the Hamiltonian is in any way valid for other systems, since the
harmonic oscillator is a notoriously pliable system.  Choosing $n = -1$ in
(\ref{51}) and restricting to $\lambda, q > 0$ produces the one-dimensional
Coulomb potential, whose associated Schr\"odinger equation can be readily
solved by standard methods.  The eigenvalue equation
\begin{equation}
\label{64}
\left ( - \frac{ \hbar^2}{ 2 m } \frac{d^2}{ dq^2} -
\frac{m}{ \lambda q} \right ) \psi_n ( q ) = E_n \psi_n ( q ) \, ,
\end{equation}
possesses the bound state energies $E_n = - m^3 / (2 \hbar^2
\lambda^2 n^2)$.  The canonically transformed Hamiltonian (\ref{51}) gives
$\beta = - 2$ for this case, so that the choice $\omega = - m^3 / ( 2
\hbar^3 \lambda^2 )$ and $\delta = 0$ reproduces the bound state energy
spectrum of the Schr\"odinger equation since $\tilde{H} = - \hbar | \omega |
( a^\dagger a + 1 )^{-2} $.  In addition, it is possible to evaluate the
integral (\ref{58}) for this case.  Following the prescription outlined in
(\ref{60}) and (\ref{61}) and using the value $\omega$ determined from the
differential equation gives the result
\begin{equation}
\label{65}
\hbar^{3/2} ( Q - Q_i ) =
- \left ( \frac{ \lambda^2 }{ 2 m^3} \right )^{\frac{1}{2}}
pq \sqrt{ \frac{m}{\lambda q} - \frac{p^2}{2m} }
- 2 \arcsin \sqrt{ \frac{ \lambda q p^2 }{ 2 m^2 } } \, .
\end{equation}
The associated annihilation operator (\ref{62}) possesses the factor
\begin{eqnarray}
\label{66}
\exp ( - i Q \hbar^{-3/2} ) & \propto &
\exp \left ( 2 i \arcsin \sqrt{ \frac{ \lambda q p^2 }{ 2 m^2 } } \right )
\nonumber \\
& = & - \frac{ 2 \lambda q}{m} \left (
H ( p , q ) + \frac{m}{ 2 \lambda q } \right )
+ 2 i \sqrt{ - \frac{\lambda^2 q^2 p^2}{ 2 m^3} H(p,q) } \; .
\end{eqnarray}
Setting the Hamiltonian equal to its ground state eigenvalue, $E_1 = - m^3 /
( 2 \lambda^2 \hbar^2 ) $, reduces (\ref{66}) to
\begin{equation}
\label{67}
a \, \propto \, \frac{m^2}{ \lambda \hbar^2} q - 1 + \frac{i}{\hbar} q p
\, \rightarrow \,
\frac{m^2}{ \lambda \hbar^2} q - 1 +  q \frac{ \partial }{ \partial q }
\, ,
\end{equation}
and this differential operator annihilates the ground state wave function
determined from the Schr\"odinger equation, $\psi_0  = C q \exp ( -
m^2 q / \hbar^2 \lambda ) $.

As another example, letting $n \rightarrow \infty$ in (\ref{51}) produces a
potential that is zero for $|q| < 1 / \lambda$ and infinite for $|q| > 1 /
\lambda$.  This limit therefore corresponds to a particle in an infinite
well of width $2 / \lambda$.  In this limit $\beta \rightarrow 2$, so that
choosing $\omega = \pi^2 \lambda^2 \hbar / 8 m$ and $\delta = -1$ allows
(\ref{56}) to reproduce the standard square well energy spectrum $E_n = n^2
\hbar \omega$.  The form for $f(Q)$ given by (\ref{58}) for this limit is
not useful since it can be shown to correspond to a mapping of the interval
$2/ \lambda$ into the whole real line.

It is, however, possible to solve the classical square well problem using a
canonical transformation of the form $F ( p , Q ) = - p \, f(Q)$, where
the function $f$ is chosen to be
\begin{equation}
\label{68}
f (  Q ) = \frac{8}{ \lambda \pi^2} \sum_{n = 1, 3, 5, \ldots}^{\infty}
\frac{ (-1)^{(n-1)/2} }{ n^2} \sin ( n \pi Q ) \, .
\end{equation}
The Fourier series (\ref{68}) is the sawtooth wave with unit period and
maxima and minima of $ \pm \lambda^{-1}$.  The derivative of (\ref{68})
gives the square wave with values $ \pm 2 \lambda^{-1}$, so that
\begin{equation}
\label{69}
\left ( \frac{ \partial f ( Q ) }{ \partial Q } \right )^2
= \frac{4}{ \lambda^2} \, .
\end{equation}
As a result, the free Hamiltonian is mapped into another free Hamiltonian
under the action of the canonical transformation,
\begin{equation}
\label{70}
P = p \, \frac{ \partial f ( Q ) }{ \partial Q} \,
\Rightarrow \,
\frac{p^2}{ 2 m } \, \rightarrow \, \frac{ \lambda^2 P^2}{ 8 m } \, ,
\end{equation}
so that the classical result for the evolution of $Q$ is
\begin{equation}
\label{71}
Q = Q_i + \frac{ \lambda^2 P t }{ 4 m } \, .
\end{equation}
Whereas the original momentum $p$ oscillates between a positive and negative
value, the new variable $P$ is truly a constant of motion.  The classical
canonical transformation gives
\begin{equation}
\label{72}
q = f \left ( Q_i + \frac{ \lambda^2 P t }{ 4 m } \right )  \, , \; \;
p = \frac{ P }{ \frac{\partial f}{ \partial Q}} \, ,
\end{equation}
and this describes the bouncing motion of the classical particle in the
square well.

\newsection{Quantum Canonical Transformations and Anomalies}

In the operator approach to quantum mechanical systems any nontrivial change
of variables is complicated by the ordering and noncommutativity of the
constituent operators that occur in expressions.  Such difficulties are not
immediately apparent in the path integral expression (\ref{seven}) due to
the $c$-number form of the variables in the action.  However, closer
inspection of the action in (\ref{seven}) shows that the formal time
derivatives do not behave in a way that allows the classical canonical
transformation to be implemented, since $q_{j+1} - q_j$ is not {\it a
priori} $O( \epsilon )$.  For this reason the implementation of canonical
transformations in the path integral formalism cannot in general reproduce the
transformed classical action.  In fact, it would be an error in most cases
if it did, since using the classical result in the action of the transformed
path integral would yield a transition element that was inconsistent with
the results obtained from the Schr\"odinger equation, operator techniques,
or the original untransformed path integral.  An alternate approach must be
taken, and in this paper a variant of the method of Fukutaka and Kashiwa
\cite{Fukutaka} will be used.  This approach can be inferred by examining the
ramifications of using a new set of canonically conjugate variables to
construct the path integral.

The phase space of a quantum mechanical system may possess unusual
properties, as the following simple argument demonstrates.  The standard
configuration space transition element can be written
\begin{equation}
\label{77.2}
\langle \, q(0) \, | \, q(T) , T \, \rangle = G ( q(0), q(T), T )
\exp i W ( q(0), q(T), T) \; ,
\end{equation}
and upon taking the modulus squared, integrating over $q(T)$, and using the
completeness of the position states, it follows that
\begin{equation}
\label{77.3}
\int dq(T) \, | G ( q(0), q(T), T) |^2 =
\langle \, q(0) \, | \, q(0) \, \rangle = \int \frac{ dp(0) }{ 2 \pi \hbar}
\; .
\end{equation}
For a quadratic Hamiltonian, it is well known that the function $G$ is
independent of $q(0)$ and $q(T)$ \cite{Feyn}, so that (\ref{77.3}) relates
the volumes of quantum phase space components to each other.  For example,
the free particle is such that
\begin{equation}
G ( q(0), q(T) , T ) = \sqrt{ \frac{m }{ 2 \pi i \hbar T } } \; ,
\end{equation}
so that (\ref{77.3}) gives
\begin{equation}
\label{77.4}
\int dq(T) = \frac{T}{m} \int dp(0) \; .
\end{equation}
Result (\ref{77.4}) is reminiscent of the spreading of a wave-packet for the
free particle.  A similar analysis for the harmonic oscillator gives
\begin{equation}
\int dq(T) = \frac{ \sin \omega T }{ m \omega } \int dp(0) \; .
\end{equation}
Of course, both of the phase space volumes appearing in these expressions
are infinite, and the comparison of infinities is a poorly defined
endeavor.  Nevertheless, these results hint at a richer structure in the
quantum mechanical phase space, and that this structure is related to the
prefactor $G$.

If there exist new conjugate operators, $\hat{Q}$ and $\hat{P}$, at the
quantum level, it is then natural to construct the path integral using their
eigenstates as intermediate states.  This means a repetition of the steps
used in Sec.~II that led to (\ref{seven}), using as a unit projection operator
\begin{equation}
\label{78}
\hat{\openone} = \int \frac{dP}{ 2 \pi \hbar } \, dQ \,
| \, Q \, \rangle e^{i Q P / \hbar } \langle \, P \, |  \, .
\end{equation}
In so doing, two difficulties occur immediately.  The first is the evaluation
of the matrix elements of the original Hamiltonian, $H ( \hat{p}, \hat{q}, t
)$, in the new states.  The second problem is the endpoint evaluation.  While
the intermediate states are the new ones, the endpoint states of the
transition element are still eigenstates of the old operators. In the
transition element of (\ref{one}) there are two inner products of importance
to the final form of the path integral constructed using $N$ copies of the
unit projection operator (\ref{78}), and these are $\langle \, p_f \, | \,
Q_N \, \rangle$ and $ \langle \, P_1 \, | \, q_i \, \rangle$.  In some
simple cases, such as the transformation from Cartesian to polar coordinates,
it is possible to obtain exact expressions for these inner products.  In
most cases it is not.  In order to evaluate these inner products, a general
form for them will be assumed, and a consistency condition necessary to
maintain (\ref{78}) as a unit projection operator will be derived.  This
result will serve to define a quantum mechanical version of canonical
transformations that is similar in structure to that proposed by Fukutaka
and Kashiwa \cite{Fukutaka}.

If (\ref{78}) is to hold, the form of the inner products must be such that
\begin{equation}
\label{79}
\langle \, p_f \, | \, q_i \, \rangle =
\int \frac{ dP_1 }{ 2 \pi \hbar } \, dQ_N \,
\langle \, p_f \, | \, Q_N \, \rangle \, e^{i P_1 Q_N / \hbar }
\langle \, P_1 \, | \, q_i \, \rangle
\, .
\end{equation}
The new variables and the inner products are defined in the following way.
The inner products are written formally in terms of some function $F ( p , Q
) $,
\begin{eqnarray}
\label{80}
\langle \, p_f \, | \, Q_N \, \rangle & = &
\exp \left \{ \frac{i}{\hbar} \left [ P_f ( Q_f - Q_N )  +
F ( p_f , Q_f ) \right ] \right \}
\, ,
\\
\label{81}
\langle \, P_1 \, | \, q_i \, \rangle & = &
\exp \left \{ - \frac{i}{\hbar} \left [  P_1  Q_i +
F ( p_i , Q_i ) \right ] \right \}
\, .
\end{eqnarray}
Inserting forms (\ref{80}) and (\ref{81}) into (\ref{79}) gives
\begin{equation}
\label{82}
\langle \, p_f \, | \, q_i \, \rangle
= \exp \left \{ \frac{i}{\hbar} \left [ P_f ( Q_f - Q_i ) +
F ( p_f , Q_f ) -
F ( p_i , Q_i ) \right ] \right \} \, .
\end{equation}
In order that (\ref{82}) reduces to the standard result, it is necessary
to identify
\begin{eqnarray}
\label{83}
- P_f ( Q_f - Q_i ) & = & F( p_f , Q_f ) - F ( p_f , Q_i ) \, , \\
\label{84}
- q_i ( p_f - p_i ) & = & F( p_f , Q_i ) - F ( p_i , Q_i ) \, .
\end{eqnarray}
For the identifications of (\ref{83}) and (\ref{84}) the inner product
of (\ref{82}) reduces to
\begin{equation}
\label{85}
\langle \, p_f \, | \, q_i \, \rangle
= \exp \left [ - \frac{i}{\hbar} \, q_i ( p_f - p_i ) \right ] \, ,
\end{equation}
which is the correct result if the restriction $p_i = 0$, familiar from
the discussion in Sec.~II.B, is enforced.

Although identifications (\ref{83}) and (\ref{84}) result in infinite
series definitions of the new variables $P$ and $Q$, the leading term of
the expansions reproduces the classical result.  For example, (\ref{83})
gives
\begin{equation}
\label{86}
P_f = - \frac{ \partial F ( p_f , Q_f )}{ \partial Q_f }
+ \frac{1}{2} \frac{ \partial^2 F( p_f, Q_f ) }{ \partial {Q_f}^2}
( Q_f - Q_i ) + \ldots \, ,
\end{equation}
so that the first term coincides with the time independent form of the
classical canonical transformation to the new variable $P$.  In addition,
the quantum counterparts of identities such as (\ref{45.1}) are altered.
This will be discussed later in this section.

The ``infinitesimal'' versions of (\ref{83}) and (\ref{84}), to be used in
defining the path integral variables, are given by
\begin{eqnarray}
\label{87}
P_j  & = & - \frac{ F( p_j , Q_j ) - F ( p_j , Q_{j-1} ) }
{ \Delta Q_j } \, , \\
\label{88}
q_j  & = & - \frac{ F( p_{j+1} , Q_j ) - F ( p_j , Q_j ) }{ \Delta p_j }\, ,
\end{eqnarray}
where $\Delta Q_j = Q_j - Q_{j-1}$ and $\Delta p_j = p_{j+1} - p_j $.  Using
these definitions of the new variables allows the formal time derivatives
in the path integral action to be transformed appropriately, since (\ref{87})
and (\ref{88}) give
\begin{equation}
\label{89}
- q_j ( p_{j+1} - p_j ) = P_{j+1} ( Q_{j+1} - Q_j ) + F ( p_{j+1} , Q_{j+1} )
- F ( p_j , Q_j ) \, ,
\end{equation}
and the action sum in the path integral (\ref{seven}) therefore becomes
\begin{equation}
\label{90}
- \sum_{j=0}^{N} q_j ( p_{j+1} - p_j ) =
F( p_f , Q_f ) - F(p_i , Q_i ) + \sum_{j=0}^{N} P_{j+1} ( Q_{j+1} - Q_j )
\, .
\end{equation}
Result (\ref{90}) is similar in form to the standard endpoint terms
generated in the action by a canonical transformation.  It is important to
remember that this result is valid only for the case that $p_i = 0$.

The form of the transformed Hamiltonian appearing in the action is
complicated by the dependence of the old variables, $q$ and $p$, on $\Delta
P$ and $\Delta Q$, as well as $P$ and $Q$.  From (\ref{87}) it follows that
$p_j$ is a function of $P_j$, $Q_j$, and $Q_{j-1}$, but that the dependence
on $Q_{j-1}$ can be expressed in a power series in $\Delta Q_j$,
\begin{equation}
\label{91}
P_j = \sum_{n=1}^{\infty} \frac{1}{n!} \frac{ \partial^n F}{ \partial
{Q_j}^n} \, ( - \Delta Q_j )^{n -1} \; .
\end{equation}
The importance of the leading behavior of $\Delta Q$ in $\epsilon$, discussed
in Sec.~II.B, is now apparent. The form of the transformed Hamiltonian will
depend critically on whether the terms containing $ \Delta Q$ and $\Delta P$
are suppressed by the overall factor of $\epsilon$ that prefaces the
Hamiltonian.  It should be clear from the discussion of Sec.~II that these
$\Delta$ terms will be suppressed by {\em some} power of $\epsilon$; it is
not clear until the specific system and transformation are chosen if they
will still contribute to the path integral when the infinite sum is
evaluated. If they do, then the transformed quantum mechanical Hamiltonian
will differ from the classical transformed Hamiltonian.  The transformed
Hamiltonian will therefore be written $ H ( p_j , q_j ) = \tilde{H} ( P_j ,
Q_j, \Delta P_j, \Delta Q_j )$, since it is not {\it a priori} obvious that
the $\Delta$ terms can be suppressed.  From the discussion in Sec.~II.B it
is apparent there are cases where such terms can contribute to the
evaluation of the path integral, and, at least for one of the cases discussed
there, can become $O( \hbar )$ terms.  These are the path integral
counterparts of the ordering ambiguities in the operator approach to
canonical transformations, and, in a loose sense, represent commutators
between the old canonical variables $\hat{q}$ and $\hat{p}$.

These results can be generalized to the case that the function $F$ or the
original Hamiltonian have explicit time-dependence.  Denoting the function
as $F ( p_j, Q_j, t_j )$, an analysis similar to that which led to (\ref{87})
and (\ref{88}) gives
\begin{eqnarray}
\label{90a}
P_j & = & - \frac{ F(p_j, Q_j, t_j) - F (p_j, Q_{j-1}, t_j )}{\Delta Q_j}
\; , \\
\label{90b}
q_j &=& - \frac{ F(p_{j+1}, Q_j, t_j) - F (p_j, Q_j, t_j )}{\Delta p_j}
\; , \\
\label{90c}
H ( p_j , q_j , t_j ) & = &
\tilde{H} ( P_j , Q_j , \Delta Q_j , \Delta P_j , t_j ) +
\frac{ \partial F ( p_{j+1} , Q_j , t_j ) }{ \partial t_j } \; .
\end{eqnarray}
The result (\ref{90c}) is valid only in the limit that $t_{j+1} - t_j =
\epsilon \rightarrow 0$. However, the identifications of (\ref{90a}) through
(\ref{90c}) lead to a result similar to (\ref{90})
\begin{eqnarray}
\label{91a}
& & \sum_{j=0}^N \left [ - q_j ( p_{j+1} - p_j ) - \epsilon
H( p_j , q_j , t_j ) \right ] =
F ( p_f , Q_f , t_f ) - F ( p_i , Q_i , t_i ) \nonumber \\
& & + \sum_{j = 0}^N
\left [ P_{j+1} ( Q_{j+1} - Q_j ) - \epsilon \tilde{H} ( P_j , Q_j ,
\Delta P_j , \Delta Q_j , t_j ) - \epsilon \frac{ \partial F ( p_{j+1} ,
Q_j , t_j )}{ \partial t_j } \right ] \; .
\end{eqnarray}

However, it will now be shown that the Jacobian of a general transformation
may contribute terms of $O ( \Delta Q)$ and $ O ( \Delta P ) $ to the action
in such a way that they are not prefaced by a factor of $\epsilon$.  For that
reason they cannot be ignored, since the sum in which they are embedded
allows them to contribute a finite amount to the transformed action.  These
$ O ( \Delta Q )$ and $ O ( \Delta P ) $ contributions are calculated from
(\ref{88}) and (\ref{89}) using the implicit dependence of $Q$ on $q$ and
$p$.  Initially, these contributions will be calculated for the case of a
one-dimensional system, and the generalization will be discussed afterward.

The starting point is the definition of the inverse Jacobian,
\begin{equation}
\label{92}
J^{-1} = \prod_{j=1}^N \left [ \frac{ \partial Q_j}{ \partial q_j }
\frac{ \partial P_j }{ \partial p_j } - \frac{ \partial P_j}{ \partial q_j }
\frac{ \partial Q_j }{ \partial p_j } \right ] \; ,
\end{equation}
where it has been assumed that $Q_j$ and $P_j$ depend primarily on $q_j$ and
$p_j$, {\it i.e.}, that the dependence on the other variables is suppressed
by some power of $\epsilon$.  It will be seen that this is a self-consistent
assumption.  The partial derivatives of $P_j$ can be obtained to $O ( \Delta
Q )$ from the expansion (\ref{90}) by using the implicit dependence of $Q_j$
on $q_j$ and $p_j$.  The result is
\begin{eqnarray}
\label{93}
\frac{\partial P_j}{ \partial p_j} & = & -
\frac{ \partial^2 F}{ \partial p_j \, \partial Q_j} -
\frac{1}{2} \frac{ \partial^2 F}{ \partial {Q_j}^2 }
\frac{ \partial Q_j }{ \partial p_j } +
\left [ \frac{1}{2} \frac{ \partial^3 F}{ \partial p_j
\, \partial {Q_j}^2 }  + \frac{1}{6} \frac{ \partial^3 F}{ \partial {Q_j}^3}
\frac{\partial Q_j}{ \partial q_j} \right ] \Delta Q_j \\
\label{94}
\frac{\partial P_j}{ \partial q_j} & = & - \frac{1}{2}
\frac{ \partial^2 F}{ \partial {Q_j}^2} \frac{\partial Q_j}{ \partial q_j}
+ \frac{1}{6} \frac{ \partial^3 F}{ \partial {Q_j}^3} \frac{ \partial
Q_j}{\partial q_j} \Delta Q_j \; .
\end{eqnarray}
Direct substitution of (\ref{93}) and (\ref{94}) into (\ref{92}) gives
\begin{equation}
\label{95}
J^{-1} = \prod_{j=1}^N \left [ - \frac{ \partial^2 F}{ \partial p_j \,
\partial Q_j } \frac{ \partial Q_j }{ \partial q_j } +
\frac{1}{2} \frac{ \partial^3 F}{ \partial p_j \, \partial {Q_j}^2 }
\frac{ \partial Q_j }{ \partial q_j } \Delta Q_j \right ] \; .
\end{equation}

Result (\ref{88}) can now be differentiated and combined with the
independence of $q_j$ and $p_j$ to obtain, to $O ( \Delta p )$, the
quantum counterpart of (\ref{45.1}),
\begin{equation}
\label{96}
1 = \frac{\partial q_j}{ \partial q_j} =
- \frac{ \partial^2 F}{ \partial p_j \, \partial Q_j } \frac{ \partial Q_j}
{\partial q_j} - \frac{1}{2} \frac{ \partial^3 F}{ \partial {p_j}^2 \,
\partial Q_j} \frac{ \partial Q_j}{ \partial q_j } \Delta p_j \; .
\end{equation}
The term $\Delta p_j$ can be written in terms of an expansion in $\Delta Q_j$
and $\Delta P_j$, so that
\begin{equation}
\label{97}
\Delta p_j = \frac{ \partial p_j}{ \partial Q_j } \Delta Q_j
+ \frac{ \partial p_j }{ \partial P_j } \Delta P_j \; .
\end{equation}
Combining (\ref{97}) and (\ref{96}) and inserting the result into (\ref{95})
yields
\begin{equation}
\label{98}
J^{-1} = \prod_{j=1}^{N} \left [ 1 +
\frac{1}{2} \left ( \frac{ \partial^3
F}{ \partial {p_j}^2 \, \partial Q_j } \frac{ \partial Q_j }{ \partial q_j}
\frac{ \partial p_j}{ \partial Q_j } + \frac{ \partial^3 F}{ \partial p_j
\, \partial {Q_j}^2 } \frac{ \partial Q_j }{\partial q_j} \right ) \Delta Q_j
+ \frac{1}{2} \frac{ \partial^3 F}{ \partial {p_j}^2 \, \partial Q_j}
\frac{ \partial Q_j}{ \partial q_j} \frac{ \partial p_j }{ \partial P_j}
\Delta P_j
\right ]
\; .
\end{equation}

In general the lack of invariance for the measure of a path integral under a
transformation, which itself is a symmetry of the action, is referred to as
an anomaly \cite{Anomaly}.  In the case of (\ref{98}), the anomaly arises
due to the formal nature of time-derivatives in the path integral action,
and has nothing to do with the behavior of the classical action under a
canonical transformation.  Nevertheless, (\ref{98}) will be referred to as
the anomaly and can be written, to lowest order, as
\begin{equation}
\label{99}
J^{-1} = \prod_{j=1}^N ( 1 + A_j \Delta Q_j + B_j \Delta P_j ) \; ,
\end{equation}
where
\begin{eqnarray}
\label{99.2}
A_j & = &
\frac{1}{2} \left ( \frac{ \partial^3
F}{ \partial {p_j}^2 \, \partial Q_j } \frac{ \partial Q_j }{ \partial q_j}
\frac{ \partial p_j}{ \partial Q_j } + \frac{ \partial^3 F}{ \partial p_j
\, \partial {Q_j}^2 } \frac{ \partial Q_j }{\partial q_j} \right )
\\
\label{99.3}
B_j & = & \frac{1}{2} \frac{ \partial^3 F}{ \partial {p_j}^2 \, \partial Q_j}
\frac{ \partial Q_j}{ \partial q_j} \frac{ \partial p_j }{ \partial P_j}
\; .
\end{eqnarray}

It is important to note that, even if $\Delta Q_j$ is $O ( \epsilon )$,
the cross-terms in (\ref{99}) can contribute finite quantities.  This follows
from the fact that
\begin{equation}
\label{100}
J^{-1} =
\lim_{N \rightarrow \infty} \prod_{j=1}^N ( 1 + A_j \Delta Q_j + B_j \Delta
P_j) = \lim_{N \rightarrow \infty} \exp \left [ \sum_{j=1}^N
\ln ( 1 + A_j \Delta Q_j + B_j \Delta P_j ) \right ] \; .
\end{equation}
As a result, the expansion of the logarithm creates terms of the form
\begin{equation}
\label{101} J = \exp \left \{ \frac{ i }{ \hbar } \sum_{j=1}^{N} \left [ i
\hbar A_j \Delta Q_j + i \hbar B_j \Delta P_j \right ] \right \}
\end{equation}
which can be absorbed into the transformed action of the path integral.  It
should be noted that these terms can contribute a finite quantity to the
action even if $\Delta Q$ is $O( \epsilon )$ since $N \epsilon \rightarrow
T$.  For the same reason, if the $\Delta$ terms are $O ( \epsilon )$ or
smaller, then the higher powers in the expansion of the logarithm can be
dropped.  Because they are proportional to $\hbar$, terms of the form
(\ref{101}) are reminiscent of the velocity-dependent potentials (\ref{22})
discussed in detail in Sec.~II.B.  Clearly, if the $\Delta$ terms are not
suppressed by a factor of $\epsilon$, it will be necessary to retain higher
order terms in both the expansion of the Jacobian (\ref{92}) as well as
later in the expansion of the logarithm in (\ref{100}).

These results may be generalized to the multidimensional case.  The
multidimensional versions of (\ref{87}) and (\ref{88}) are given by
\begin{eqnarray}
\label{102}
- P^a_j  \Delta Q^a_j & = & F ( p^a_j , Q^a_j ) - F ( p^a_j , Q^a_{j-1} )
\; , \\
\label{103}
- q^a_j  \Delta p^a_j & = & F ( p^a_{j+1} , Q^a_j ) - F ( p^a_j , Q^a_j )
\; ,
\end{eqnarray}
where a sum over the repeated index $a$ is implicit.  The definitions
(\ref{102}) and (\ref{103}) do not yield a unique expression for each
of the $q^a_j$ and $P^a_j$ since the Taylor series expansions can be
separated in an arbitrary manner for each of the variables.  In what follows
a symmetrized definition of each of the canonical variables will be used,
so that
\begin{eqnarray}
\label{104}
P^a_j & = & - \frac{ \partial }{ \partial Q^a_j } \sum_{n=0}^{\infty}
\frac{(-1)^n}{(n+1)!} \frac{\partial^n F_j }{\partial {Q^{a_1}_j} \cdots
\partial {Q^{a_n}_j}} \Delta Q^{a_1}_j \cdots \Delta Q^{a_n}_j \; , \\
\label{105}
q^a_j & = & - \frac{ \partial }{ \partial p^a_j} \sum_{n=0}^{\infty}
\frac{(-1)^n}{(n+1)!} \frac{\partial^n F_j }{\partial {p^{a_1}_j} \cdots
\partial {p^{a_n}_j}} \Delta p^{a_1}_j \cdots \Delta p^{a_n}_j \; ,
\end{eqnarray}
where there is an implicit sum over any repeated pair of $a_i$ coordinate
indices. It is straightforward to repeat the analysis that led to (\ref{99})
and (\ref{99.2}) and this yields the multidimensional version of the anomaly
to $O( \Delta )$,
\begin{equation}
\label{106}
J^{-1} = \prod_{j=1}^N ( 1 + A^a_j \Delta Q^a_j + B^a_j \Delta P^a_j ) \; ,
\end{equation}
where
\begin{eqnarray}
\label{107}
A^a_j & = & \frac{1}{2}
\frac{ \partial^3 F_j }{ \partial p^b_j \,
\partial p^c_j \, \partial Q^d_j } \, \frac{ \partial p^b_j}{ \partial Q^a_j}
\, \frac{ \partial Q^d_j}{ \partial p^c_j} +
\frac{1}{2}
\frac{ \partial^3 F_j }{ \partial p^b_j \,
\partial Q^c_j \, \partial Q^a_j } \, \frac{ \partial Q^b_j}{ \partial q^c_j}
\; , \\
\label{108}
B^a_j & = & \frac{1}{2}
\frac{ \partial^3 F_j }{ \partial p^b_j \,
\partial p^c_j \, \partial Q^d_j } \, \frac{ \partial p^b_j}{ \partial P^a_j}
\, \frac{ \partial Q^d_j}{ \partial q^c_j} \; .
\end{eqnarray}
Exponentiation of (\ref{106}) leads to a result similar to (\ref{101}).

The $O ( \Delta )$ anomaly takes a particularly simple form when the
original generating function is given, for the one-dimensional case, by
\begin{equation}
\label{109}
F = - p^\alpha f ( Q ) \; .
\end{equation}
{}From the results of Sec.~III it is clear that (\ref{109}) is adequate to
transform all arbitrary single power potentials to a cyclic form.  The
anomaly associated with (\ref{109}) will be evaluated using the classical
forms for the new variables.  Such a procedure is consistent only to $O (
\Delta )$.  It follows that these classical forms are given by solving
\begin{equation}
\label{110}
q_j = \alpha ({p_j})^{( \alpha - 1 )} f( Q_j )
\; , \; \;
P_j = ({p_j})^\alpha \frac{ \partial f}{ \partial Q_j} \; ,
\end{equation}
and these relations in turn show that
\begin{eqnarray}
\label{111}
\frac{ \partial Q_j }{ \partial q_j } & = & \frac{1}{ \alpha}
{p_j}^{( 1 - \alpha ) }
\left ( \frac{ \partial f }{ \partial Q_j} \right )^{-1}  \; , \\
\label{112}
\frac{ \partial p_j }{ \partial Q_j } & = & - \frac{1}{ \alpha } {p_j}^{1 -
\alpha}
\left ( \frac{ \partial f}{ \partial Q_j} \right )^{-2}
\frac{ \partial^2 f}{ \partial {Q_j}^2 } \; .
\end{eqnarray}
Using (\ref{110}) through (\ref{112}) in (\ref{99.2}) and (\ref{99.3})
yields
\begin{eqnarray}
\label{113}
A_j & = & - \frac{1}{2 \alpha} \left ( \frac{ \partial f}{ \partial Q_j }
\right )^{-1} \frac{ \partial^2 f}{ \partial {Q_j}^2 } \; , \\
\label{114}
B_j & = & \frac{ 1 - \alpha }{ 2 \alpha P_j} \; .
\end{eqnarray}
Similarly, using the multi-dimensional generating function
\begin{equation}
F = - p^a f^a ( Q )
\end{equation}
results in a vector anomaly solely of the $A$ type, given by
\begin{equation}
\label{115a}
A^a_j = - \frac{1}{2} \frac{ \partial^2 f^b }{ \partial Q^a_j \,
\partial Q^c_j } \, \frac{ \partial Q^c_j }{ \partial q^b_j }
\end{equation}

It is important to note that if it is possible to treat $\Delta Q \approx
\epsilon \, \dot{Q}$, then the exponentiated anomaly term of (\ref{113})
becomes
\begin{equation}
\label{115}
- \lim_{N \rightarrow \infty} \sum_{j = 1}^N A_j \Delta Q_j =
- \int_0^T {\rm dt} \, A ( Q ) \, \dot{Q} =
\frac{1}{2 \alpha}
\int_0^T {\rm dt \,
\frac{d}{ dt } } \ln \frac{ \partial f(Q) }{ \partial Q} \; .
\end{equation}
For these conditions the entire $A$ anomaly therefore reduces to a prefactor
for the path integral, given by
\begin{equation}
\label{116}
A_p = \left ( \frac{ \partial f(Q_f) }{ \partial Q_f } \right )^{1/2\alpha}
\left (
\frac{ \partial f( Q_i ) }{ \partial Q_i } \right )^{- 1/2\alpha} \; .
\end{equation}
Similarly, the $B$ anomaly can be written
\begin{equation}
- \lim_{N \rightarrow \infty} \sum_{j = 1}^N B_j \Delta P_j =
- \frac{1 - \alpha}{ 2 \alpha } \int_0^T {\rm dt} \, \frac{\dot{P}}{P} =
- \frac{1 - \alpha }{2 \alpha}
\int_0^T {\rm dt \, \frac{d}{ dt } } \ln P \; .
\end{equation}
As a result, the $B$ anomaly creates a second prefactor,
\begin{equation}
\label{116a}
B_p = \left ( \frac{P_i}{P_f} \right )^{(1-\alpha)/2\alpha} \; .
\end{equation}
Results (\ref{116}) and (\ref{116a}) show that, even in the case that the
canonically transformed Hamiltonian is cyclic and the transformed path
integral generates no prefactor, it is still possible for the correct
prefactor or van Vleck determinant to be recovered from the anomaly
associated with the canonical transformation.

However, results (\ref{116}) and (\ref{116a}) also show that the problem of
identifying the appropriate boundary conditions for $Q$ and $P$ is of
paramount importance to evaluating the anomaly and determining the correct
prefactor for the original path integral.  In previous sections it has been
stressed that the use of a canonical transformation requires suppressing the
$p_i$ term that must be inserted into the action to allow the definition of
the canonical transformation.  On the face of it, simply setting $p_i$ to
zero would appear to be sufficient to bypass this problem.  However, doing
so would create three initial and final conditions for the classical system,
thereby overspecifying the classical solution to the equations of motion, a
solution that is critical to evaluating the path integral for cyclic
coordinates.  However, if $q_i$ is set to zero, the $p_i$ term is
automatically suppressed since it appears in the action as $q_i p_i$. This
choice therefore allows the value of $p_i$ to be determined from the
classical equations of motion consistent with the boundary conditions $q_i =
0$ and $p_f$ arbitrary.  The requirement that $q_i$, rather than $p_f$, be
zero for consistency is an outgrowth of choosing to write the action with
a term of the form $q \dot{p}$, rather than $p \dot{q}$. This in turn was a
result of choosing a canonical transformation of the third kind.  Other
choices will lead to different consistency requirements.

In the case of quantized variables, the problem is yet more subtle.  In
Sec.~II.D the path integral with an action translated by a classical
solution was evaluated and the fluctuation variables $p_j$ and $q_j$, given
by (\ref{33}) and (\ref{33.2}), were shown to be arbitrary at their
undefined endpoint values, {\it i.e.}, $q(t = T)$ and $p( t = 0)$.  While
this is a natural consequence of the uncertainty principle, it means that the
original quantum variables do not collapse to their classical values at
these times, {\it i.e.}, $q ( t = T ) \neq q_c ( t = T )$.  Therefore, using
the classical definitions for both of the $q$ and $p$ endpoint values is not
a reliable method.  As in the classical case, if $p_f$ is to be defined and
$p_i$ is to be arbitrary, {\it i.e.}, non-zero, it is clear from the
discussion in Sec.~II.B that the path integral must be evaluated at $q_i =
0$, since such a choice will suppress the $p_i$ term while still allowing
$p_i$ to be arbitrary.  The absence of $q_f$ from the action of the path
integral of the form (\ref{seven}) allows it to be arbitrary without
encountering a similar problem.  Thus, the canonically transformed path
integral's endpoint values are correct only if $q_i = 0$.  For a canonical
transformation of the form given by (\ref{109}), this means that $Q_i$ must
be a root of $f(Q)$.  This clearly also suppresses the initial value of the
generating function $(p_i)^\alpha f(Q_i)$.

Obviously, the $q_i \neq 0$ case can be evaluated by first translating the
action everywhere by the classical solutions, as in Sec.~II.D.  This leaves
a path integral with the effective boundary conditions $q_i = 0$ and $p_f =
0$, allowing a consistent evaluation.  A drawback to this technique is that
such a translation will create additional terms in the potential in most
cases, and the simple canonical transformations introduced in Sec.~III to
render power potentials cyclic will no longer be applicable after the
translation.  However, if the original potential was linear or quadratic
this will not be the case, since such a translation induces no additional
terms in the fluctuation potential for these two cases.  A translation by a
classical solution then shows that the prefactor of the form (\ref{116})
must be independent of the endpoint values for the case that the original
potential was linear or quadratic, and should be evaluated consistent with
the conditions $q_i = 0$ and $p_f = 0$.

Apart from these considerations, the transformed action with the anomaly
term in it is given by
\begin{equation}
\sum_{j=1}^N \left [ ( P_j + i \hbar A_j ) \Delta Q_j +
i \hbar B_j \Delta P_j - \epsilon H ( P_j , Q_j , \Delta Q_j , \Delta P_j )
\right ] \; .
\end{equation}
If the range of the $P_j$ integrations is $- \infty$ to $ + \infty$, it is
possible to move the anomaly into the Hamiltonian by translating the $P_j$
variables to $P_j - i \hbar A_j$, so that the Hamiltonian becomes formally
similar to that of a particle moving in a complex vector potential.

The anomaly appears because of the structure of quantum mechanical phase
space.  The exact function of the anomaly depends on the specific system
being evaluated.  Some of these will be discussed in Sec.~V.

\newsection{Examples}

In this section the machinery developed in the previous sections will be
applied to specific cases to evaluate the path integral by a canonical
transformation.  In most of the cases the exact form of the path integral
is available by other methods, so that the outcome of the canonical
transformation may be compared to show that equivalent results are obtained.

\subsection{ Transformations of the Free Particle }

In this subsection a specific set of canonical transformations of free
particle systems will be considered.  In Sec.~II.C the path integral
(\ref{31}) for the square well was derived.  Through Poisson resummation it
was shown to possess the same infinite range of integrations for the measure
as that of a free particle.  The path integral for the square well can
therefore be evaluated by the techniques of (\ref{24}) and (\ref{25}) for
cyclic Hamiltonians. This shows that the square well path integral reduces
to the correct result, {\it i.e.}, the value of the action along the
classical trajectory with the additional overall factor of $1 / \sqrt{2a}$.
There is no need to perform a canonical transformation on this system.

However, since the exact solution of the free particle path integral is
available, such a system can serve as a laboratory to investigate the
validity of the techniques derived in previous sections.  To begin with, the
variables in the action will be translated by the classical solution to the
equation of motion, so that the endpoint variables are given by $p_{N+1} =
0$ and $q_0 = 0$.  Because it is quadratic in the momentum, the action is
unaffected in form by this translation.  However, the arguments of Sec.~II.C
show that the remaining path integral should reduce to a factor of unity,
even in the event that it is canonically transformed.  In this subsection
the effect of canonical transformations associated with the classical
generating function $F = - p \, f( Q )$ on such a free particle path
integral will be considered.  Such a canonical transformation at the
classical level creates a Hamiltonian that, for most choices of $f$, is
velocity-dependent.  Such Hamiltonians are typically not self-adjoint,
creating difficulties in constructing the Hilbert space of the theory.  It
is therefore of interest to examine how the transformed path integral
sidesteps this problem.

This canonical transformation has the general form (\ref{109}),
so that, to $O(\Delta)$, the anomaly is given by
\begin{equation}
\label{117}
A_j = - \frac{1}{2} \left ( \frac{ \partial f }{ \partial Q_j } \right
)^{-1} \frac{ \partial^2 f }{ \partial {Q_j}^2 } \; ,
\; \; B_j = 0 \; ,
\end{equation}
It is important to investigate if the approximations used to derive
(\ref{117}) are valid, since the exact Jacobian may contain additional
terms.  The definitions of the new quantum mechanical variables in
(\ref{87}) and (\ref{88}) result in
\begin{eqnarray}
\label{118}
q_j & = & f ( Q_j ) \\
\label{119}
p_j & = & \frac{ P_j \Delta Q_j }{ f(Q_j) - f(Q_{j-1})} =
\left ( \frac{ \partial f ( Q_j ) }{ \partial Q_j }
- \frac{1}{2} \frac{ \partial^2 f ( Q_j ) }{ \partial {Q_j}^2 } \Delta Q_j
+ \ldots \right )^{-1} P_j \; .
\end{eqnarray}
Because $q_j$ is independent of $P_j$, the exact Jacobian for the $j$th
product in the measure is given by
\begin{equation}
\label{120}
dp_j \, dq_j = dP_j \, dQ_j \, J_j =
dP_j \, dQ_j \left [ 1 - \frac{1}{2}
\left ( \frac{ \partial f ( Q_j ) }{ \partial
Q_j } \right )^{-1} \frac{ \partial^2 f ( Q_j ) }{ \partial {Q_j}^2 }
\Delta Q_j  + \ldots \right ]^{-1} \; .
\end{equation}
When exponentiated, (\ref{120}) yields the same $O(\Delta)$ result as
(\ref{117}).

However, it would be misleading to exponentiate this Jacobian for the
following reason.  The Hamiltonian in the path integral action remains
quadratic in momentum, since
\begin{equation}
\label{121}
\epsilon \, \frac{ {p_j}^2 }{ 2 m } =
\epsilon \, \frac{ {P_j}^2 }{2  m } \left ( \frac{ \partial f }{ \partial
Q_j } \right )^{-2}
\left [ 1 - \frac{1}{2}
\left ( \frac{ \partial f ( Q_j ) }{ \partial
Q_j } \right )^{-1} \frac{ \partial^2 f ( Q_j ) }{ \partial {Q_j}^2 }
\Delta Q_j  + \ldots \right ]^{-2} \; .
\end{equation}
Even though the $O(\Delta Q)$ terms in the Hamiltonian could be treated as a
perturbation, the presence of the $\Delta Q$ terms in the anomaly prevent
integrating over the $Q$ variables as in (\ref{24}) to show that this
remaining path integral reduces to unity.  Instead, the $P$ integrations
must be performed first, and this shows that the anomaly in the measure is
cancelled as a result of the Gaussian $P$ integrations.  Since the action
was translated by the classical solution {\em prior}\/ to canonical
transformation, the boundary conditions are $P_f = P_i = 0$ and $Q_f = Q_i =
0$.  Upon performing the $P$ integrations, the remaining Euclidean path
integral reduces to
\begin{eqnarray}
\label{122} & & \int \prod_{i=1}^N \left [ dQ_i \, \frac{\partial f (Q_i )
}{ \partial Q_i} \sqrt{ \frac{ 1 }{  2 \pi \hbar \epsilon }} \right ]
\nonumber \times \\
& & \exp \left \{ - \frac{1}{\hbar}
\sum_{j=1}^N \frac{  m \Delta {Q_j}^2 }{ 2 \epsilon } \left ( \frac{ \partial
f ( Q_j ) }{ \partial Q_j } \right )^2
\left [ 1 - \frac{1}{2}
\left ( \frac{ \partial f ( Q_j ) }{ \partial
Q_j } \right )^{-1} \frac{ \partial^2 f ( Q_j ) }{ \partial {Q_j}^2 }
\Delta Q_j  + \ldots \right ]^{2} \right \} \; .
\end{eqnarray}
It is natural to define the new variables $f_j = f( Q_j)$, and this gives
\begin{equation}
df_i = dQ_i \frac{ \partial f (Q_i) }{ \partial Q_i } \; .
\end{equation}
This new variable must have the same range of integration as the original
variable $q_j$ by virtue of (\ref{118}).  The transformed action simplifies
as well since
\begin{eqnarray}
\label{123}
& & \frac{  m \Delta {Q_j}^2 }{ 2 \epsilon }
\left ( \frac{ \partial
f ( Q_j ) }{ \partial Q_j } \right )^2
\left [ 1 - \frac{1}{2}
\left ( \frac{ \partial f ( Q_j ) }{ \partial
Q_j } \right )^{-1} \frac{ \partial^2 f ( Q_j ) }{ \partial {Q_j}^2 }
\Delta Q_j  + \ldots \right ]^{2} \nonumber \\
& & =
\frac{  m \Delta {Q_j}^2 }{ 2 \epsilon }
\left ( \frac{ \Delta f_j }{ \Delta Q_j } \right )^2 =
\frac{  m \Delta {f_j}^2 }{ 2 \epsilon } \; .
\end{eqnarray}
The resulting path integral is therefore identical to the original
path integral written in terms of the $q_j$ variables.  The anomaly has
been cancelled by contributions from the Hamiltonian.  This means that the
path integral defined by the measure (\ref{120}) and the action (\ref{121})
maintains a well-defined quantum theory for a velocity-dependent Hamiltonian.

In general, it is not difficult to see that a canonical transformation
resulting in a transformed Hamiltonian that is quadratic in $P$ will possess
$O ( \Delta Q ) $ terms that, upon integration of the $P$ variables, can
result in cancellation of the anomaly.

\subsection{ The Linear Potential }

The case of the linear potential,
\begin{equation}
\label{125}
H = \frac{p^2}{2 m} + m \lambda q \; ,
\end{equation}
allows an exact integration of the path integral, yielding the transition
element
\begin{equation}
\label{126}
W_{fi} = \frac{1}{\sqrt{2 \pi \hbar}} \exp \left \{
- \frac{ i }{ \hbar } \left [
\frac{1}{2} T^2 \lambda p_f + \frac{ T {p_f}^2}{ 2m }
+ q_i m \lambda T + p_f q_i + \frac{1}{6} m \lambda^2 T^3 \right ]
\right \} \; .
\end{equation}
Since the action is linear, the effect of a canonical transformation on the
path integral will be analyzed for the case that $p_f = q_i = 0$.  Result
(\ref{126}) shows that the path integral with $p_f = q_i = 0$ must result in
\begin{equation}
\label{127}
W_{fi} = \frac{1}{\sqrt{2 \pi \hbar}} \exp \left \{
- \frac{ i }{ \hbar }
\frac{1}{6} m \lambda^2 T^3
\right \} \; .
\end{equation}

The evaluation of this path integral by canonical transformation can be used
as another test of the techniques developed in the previous sections.  The
classical action has the form (\ref{51}) and can be rendered cyclic by a
canonical transformation of the type (\ref{52}).  Evaluating the integral
(\ref{58}) for the classical generating function yields
\begin{equation}
\label{128}
F(p, Q) = - \frac{ p^3 }{ 6 m^2 \lambda } \left [ \frac{ 8 }{ 9 m \lambda^2
Q^2 } - 1 \right ] \; .
\end{equation}
However, this generating function suffers from a defect inherited from the
parent Hamiltonian, which is not positive-definite due to the odd power of
$q$.  Using the generating function of (\ref{128}) yields the classical
Hamiltonian
\begin{equation}
\label{129}
\tilde{H} = P^{2/3} \; ,
\end{equation}
which is positive-definite since $P$ is assumed to range over real values
and the real branch of the 2/3 power is used.  In order to match the range
of the original Hamiltonian, $P$ would have to range over both pure real and
pure imaginary values, rendering the integrations over $P$ undefined.  A
similar problem exists for the range of the new canonical variable $Q$,
since classically it is transformed to
\begin{equation}
\label{130}
Q^2 = \frac{ 8 }{ 9 m \lambda^2 } \frac{ p^2 }{ p^2 + 2 m^2 \lambda q}
\; ,
 \end{equation}
resulting in imaginary values for the case that the original Hamiltonian is
negative.

This problem can be remedied by adding the term $p E_0 / m \lambda$ to the
generating function (\ref{128}), where the limit $E_0 \rightarrow \infty$ is
understood.  Doing so allows the range for $P$ to be real while still
matching the range of the original Hamiltonian, since the transformed
Hamiltonian becomes
\begin{equation}
\label{131}
\tilde{H} = P^{2/3} - E_0 \; \Rightarrow \; P = \left (
\frac{p^2}{ 2m } + m \lambda q + E_0 \right )^{3/2} \; ,
\end{equation}
while the range of $Q$ is now real, since
\begin{equation}
\label{132}
Q^2 = \frac{ 8 }{ 9 m \lambda^2 } \frac{ p^2 }{ ( p^2 + 2 m^2 \lambda q +
2 m E_0 ) }
\; .
\end{equation}
The necessary presence of $E_0$ stems from the fact that the Hamiltonian is
not bounded from below.

Since the transformation does not yield a quadratic Hamiltonian, it will be
assumed that the perturbative argument of Sec.~II is valid, and that terms
of $O(\Delta Q)$ in the transformed Hamiltonian can be suppressed.  The
transformed path integral is then given by
\begin{equation}
\label{133}
W_{fi} =  \frac{A_p B_p}{ \sqrt{2 \pi \hbar} }
\exp \left \{ \frac{ i }{ \hbar } [ F_f - F_i ] \right \}
\int \frac{ dP}{ 2 \pi \hbar } \, dQ \, \exp \left \{ \frac{i}{ \hbar }
\int_0^T {\rm dt} \, [ P \dot{Q} - P^{2/3} + E_0 ] \right \}
\; ,
\end{equation}
where $A_p$ and $B_p$ are the anomaly prefactors (\ref{116}) and
(\ref{116a}), $F_i$ and $F_f$ are the generating function evaluated at the
initial and final conditions, and all $O(\Delta Q)$ terms have been
suppressed in the Hamiltonian.

Because the transformed Hamiltonian is cyclic, the results of Sec.~II.C show
that the remaining path integral can be evaluated by finding the action
along the classical trajectory.  The initial and final conditions are
determined from the equations of motion for the original variables, with the
boundary conditions that $p_f$ and $q_i$ both vanish.  The solutions for $p$
and $q$ consistent with these conditions are easily found, with the result
that $p_i = m \lambda T$.  Using (\ref{131}) then gives
\begin{equation}
\label{134}
P_i  = \left ( \frac{1}{2} m \lambda^2 T^2 + E_0 \right )^{3/2}
\; ,
\end{equation}
while
\begin{equation}
\label{135}
Q_i = \frac{2}{3} \sqrt{ \frac{T^2}{ E_0 - \frac{1}{2} m \lambda^2 T^2 } }
\; .
\end{equation}
The Hamiltonian equations of motion give
\begin{eqnarray}
\label{136}
\dot{P} & = & 0 \; \Rightarrow \; P_f = P_i \; , \\
\label{137}
\dot{Q} & = & \frac{2}{3} P^{-1/3} \; \Rightarrow \;
Q_f = Q_i + \frac{2}{3} {P_f}^{-1/3} T \; .
\end{eqnarray}
In the limit that $E_0 \rightarrow \infty$, it follows that $Q_f = Q_i$.

Using these results, the action along the classical trajectory becomes
\begin{equation}
\label{138}
\int_0^T {\rm dt} \, [ P\dot{Q} - P^{2/3} + E_0 ] =
E_0 T - \int_0^T {\rm dt} \, \frac{1}{3} {P_i}^{2/3} =
\frac{2}{3} E_0 T - \frac{1}{6} m \lambda^2 T^3 \; .
\end{equation}
The generating functions reduce to
\begin{eqnarray}
\label{139}
F_f & = & 0 \\
\label{140}
F_i & = & \frac{2}{3} E_0 T \; .
\end{eqnarray}
Finally, form (\ref{116}) for the anomaly prefactor reduces to
\begin{equation}
\label{141}
A_p = \sqrt{ \frac{ Q_i }{ Q_f} } \; \Rightarrow \;
\lim_{E_0 \rightarrow \infty} A_p = 1 \; ,
\end{equation}
while the prefactor (\ref{116a}) becomes
\begin{equation}
\label{141a}
B_p = \left ( \frac{P_f}{P_i} \right )^{1/3} = 1 \; .
\end{equation}
Combining results (\ref{138} -- \ref{141a}) gives the correct result
\begin{equation}
\langle \, p_f = 0 \, | e^{- i H T / \hbar} | \, q_i = 0 \, \rangle
= \frac{1}{ \sqrt{ 2 \pi \hbar } } \exp \left \{ - \frac{i}{ \hbar }
\frac{1}{6} m \lambda^2 T^3 \right \} \; ,
\end{equation}
showing that all reference to $E_0$ has disappeared from the problem.  It is
not difficult to extend the same analysis to the case that $q_i = 0$ and
$p_f \neq 0$ to show that the correct results follow from the canonical
transformation.

\subsection{Polar Coordinates}

The transformation from Cartesian to polar coordinates served as the first
indication that adopting classical canonical transformations to the path
integral was more complicated than expected \cite{Edwards}.  In effect, it
is a multi-dimensional version of the transformation to a velocity-dependent
potential analyzed in Sec.~V.A.  As a result, a mechanism similar to
(\ref{123}) should occur, allowing the canonically transformed path integral
to maintain its equivalence to the original path integral.

The starting point is the two-dimensional Hamiltonian
\begin{equation}
\label{142}
H = \frac{1}{2m} ( {p_x}^2 + {p_y}^2 ) \; .
\end{equation}
The action associated with this Hamiltonian may be transformed into polar
coordinates by using the classical generating function
\begin{equation}
\label{143}
F = - p_x r \cos \theta - p_y r \sin \theta \; .
\end{equation}
Of course, the quantum mechanical version of this transformation results in
terms of $O(\Delta)$ and higher.  In the following analysis, the $O(\Delta)$
terms will be retained to construct the form of the path integral under this
transformation.  It is to be remembered throughout that this is a shorthand
for the full canonical transformation.  For simplicity, the boundary
conditions will match those for the case that the original Cartesian action
has been translated by the classical solutions to the equations of motion,
so that $p_{xf} = p_{yf} = x_i = y_i = 0$.  For such a choice the remaining
path integral must reduce to a factor of unity.

The procedure is tedious but straightforward.  The new momenta and
coordinates are given by
\begin{eqnarray}
\label{144}
P_{rj} & = &
p_{xj} \left ( \cos \theta_j + \frac{1}{2} \sin \theta_j \, \Delta
\theta_j \right )
+ p_{yj} \left ( \sin \theta_j - \frac{1}{2} \cos \theta_j \, \Delta \theta_j
\right )  \; , \\
\label{145}
P_{\theta j} & = &
- p_{xj} \left ( r_j \sin \theta_j - \frac{1}{2} \sin \theta_j
\, \Delta r_j - \frac{1}{2} r_j \cos \theta_j \, \Delta \theta_j \right )
\nonumber \\
& & +
p_{yj} \left ( r_j \cos \theta_j + \frac{1}{2} \cos \theta_j \, \Delta r_j
- \frac{1}{2} r_j \sin \theta_j \, \Delta \theta_j \right ) \; , \\
\label{146}
x_j & = & r_j \cos \theta_j \; , \\
\label{147}
y_j & = & r_j \sin \theta_j \; .
\end{eqnarray}
These definitions yield $p_x$ and $p_y$ in terms of the new variables.
Substituting them into the Hamiltonian gives the transformed Hamiltonian to
$O(\Delta)$:
\begin{equation}
\label{148}
\tilde{H}_j = \frac{1}{2m} \left ( {P_{rj}}^2 +
\frac{1}{{r_j}^2} \left ( 1 - \frac{1}{2}
\frac{ \Delta r_j}{ r_j} \right )^{-2}
{P_{\theta j}}^2 \right ) \; .
\end{equation}
Retention of the $O(\Delta)$ terms is essential since the transformed
Hamiltonian (\ref{148}) is not cyclic and also remains quadratic in the
momenta.  The anomaly term can be calculated to $O(\Delta)$ directly from
the form of the transformations, or by using the multi-dimensional form
(\ref{107}). The resulting measure for the path integral transforms
according to
\begin{equation}
\label{149}
dx_j \, dy_j \, dp_{xj} \, dp_{yj} \, \rightarrow \,
d \theta_j \, dr_j \, dP_{\theta j} \, dP_{rj}
\, \left ( 1 - \frac{1}{2} \frac{\Delta r_j}{r_j} \right )^{-1} \; .
\end{equation}

It is possible to exponentiate the anomaly, resulting in terms in the
transformed action with the form
\begin{equation}
\label{150}
\left ( P_{rj} - \frac{i \hbar}{2 r_j}  \right ) \Delta r_j \; .
\end{equation}
Since the range of the $P_{rj}$ integrations is infinite, this extra term
can be transferred to the Hamiltonian by translating the $P_{rj}$ variables.
This results in
\begin{equation}
\label{151}
\frac{1}{2m} {P_{rj}}^2
\, \rightarrow \,
\frac{1}{2m} {P_{rj}}^2
+ \frac{i \hbar}{2m r_j } P_{rj}
- \frac{\hbar^2}{ 8 m {r_j}^2} \; .
\end{equation}
This is precisely the self-adjoint form (\ref{SC2}) for the Weyl-ordered
Hamiltonian in spherical coordinates discussed in Sec.~III.

However, as in the case of the velocity-dependent transformation discussed
in this section, it is misleading to exponentiate the anomaly term.  This is
demonstrated by performing the momentum integrations.  The integration over
$P_\theta$ exactly cancels the anomaly, and the resulting measure in the
path integral is $r_j \, dr_j \, d \theta_j \, ( 2 \pi / m \epsilon) $,
while the action becomes
\begin{equation}
\label{152}
\sum_{j=1}^N \left ( \frac{m}{2 \epsilon} \Delta {r_j}^2  +
\frac{m}{2 \epsilon} {r_j}^2 \left ( 1 - \frac{1}{2}
\frac{ \Delta r_j }{ r_j } \right )^2 \Delta {\theta_j}^2 \right ) \; .
\end{equation}
Using (\ref{146}) and (\ref{147}) it is straightforward to show that
(\ref{152}) is, to $O(\Delta)$, the same as
\begin{equation}
\label{153}
\sum_{j=1}^N \left ( \frac{m}{2 \epsilon} {\Delta x_j}^2
+ \frac{m}{2 \epsilon} {\Delta y_j}^2 \right ) \; ,
\end{equation}
while the measure is the same as $dx_j \, dy_j ( 2 \pi / m \epsilon)$.
Thus, the path integral with $P_r$ and $P_\theta$ integrated generates a
path integral and measure exactly equivalent to the original path integral
with $p_x$ and $p_y$ integrated.  Since the original path integral reduces
to a factor of unity, this completes the proof that the path integral with
its action constructed using (\ref{148}) and measure given by (\ref{149})
reduces to a factor of unity.  This is an $O(\Delta)$ proof of equivalence,
similar to the all order proof for the transformation to a
velocity-dependent potential discussed in Sec.~V.A. This is in effect
nothing more than a multi-dimensional version of the relationship
(\ref{123}), and could be extended to an all orders proof.

\subsection{ The Harmonic Oscillator}

The harmonic oscillator has been analyzed by employing the canonical
transformation (\ref{48})
\begin{equation}
F = - \frac{p^2}{2 m \omega} \tan Q \; ,
\end{equation}
so that, in the nomenclature of Sec.~III, $f(Q) = \tan Q / ( 2 m \omega )$
and $\alpha = 2$.  It will be reviewed here for the sake of completeness and
because certain results will be used in Sec.~V.E.  The results for the
quantum version give
\begin{eqnarray}
\label{154}
q_j & = & \frac{p_{j+1} + p_j }{ 2 m \omega } \tan Q_j \\
\label{155}
P_j \Delta Q_j & = & \frac{ {p_j}^2 }{ 2 m \omega } \left ( \tan Q_j
- \tan Q_{j-1} \right )
\end{eqnarray}
The classical canonical transformation leads to the transformed Hamiltonian
$\tilde{H} = \omega P$.  The quantum version of the transformation, given by
(\ref{154}) and (\ref{155}), results in terms of $O(\Delta)$ in the
transformed Hamiltonian.  However, because the transformed Hamiltonian is
not quadratic in $P$ and is cyclic, it will be assumed that suppressing
these terms is allowed by the perturbative argument of Sec.~II.B.  A mild
difference occurs since the range of the $P$ variable is $[ 0 , \infty ]$.
This prevents the transfer of the anomaly into the Hamiltonian.  As a
result, the anomaly terms will be evaluated using (\ref{116}) and
(\ref{116a}).

Performing the path integral using results (\ref{24}) yields the transition
element
\begin{equation}
\label{156}
W_{fi} = \frac{A_p B_p}{ \sqrt{2 \pi \hbar} }
\exp \left \{ \frac{i}{\hbar} [ F_f - F_i + S_{cl}] \right \}
\end{equation}
where $S_{cl}$ is the transformed action evaluated along a classical
trajectory,
\begin{equation}
\label{157}
S_{cl} = \int_0^T {\rm dt} \, \left ( P_c \dot{Q}_c - \omega P_c \right ) \; .
\end{equation}

Hamilton's equations of motion, $\dot{Q} = \omega$ and $\dot{P} = 0$, have
the solutions $Q_f = Q_i + \omega T$ and $P_f = P_i$, showing that $S_{cl} =
0$.  The restriction to $q_i = 0$ is satisfied by the choice $Q_i = 0$.
Using these results in (\ref{116}) and (\ref{116a}) gives the anomalies
\begin{eqnarray}
\label{158}
A_p & = & \left ( \frac{ \partial f(Q_f) }{ \partial Q_f } \right )^{1/2\alpha}
\left (
\frac{ \partial f( Q_i ) }{ \partial Q_i } \right )^{- 1/2\alpha} =
\frac{1}{ \sqrt{\cos \omega T} } \; , \\
\label{159}
B_p & = & \left ( \frac{P_i}{P_f} \right )^{(1-\alpha)/2\alpha} = 1 \; .
\end{eqnarray}
The product of the anomalies reproduces the correct prefactor (\ref{40}).
The generating functions become $F_i =  0$ and
\begin{equation}
\label{160}
F_f = - \frac{ {p_f}^2 }{ 2 m \omega } \tan \omega T \; .
\end{equation}
Comparison with (\ref{39}) and (\ref{40}) shows that combining these results
in (\ref{156}) yields the correct harmonic oscillator transition element for
the case $q_i = 0$.

\subsection{ The Time-Dependent Harmonic Oscillator}

One of the drawbacks to the techniques developed in this paper has been the
restriction $q_i = 0$.  Of course, it is possible to circumvent this problem
by first translating the action by a classical solution to the equations of
motion.  The remaining path integral will then have the boundary condition
$q_i = 0$ automatically.  Unfortunately, for all but the quadratic and linear
potentials, doing so induces additional terms into the action, preventing the
use of the generating function (\ref{52}) which was derived to render the
simple power potential potential of (\ref{51}) cyclic.

However, it is possible to treat any translated action with a potential
involving terms higher than quadratic in first approximation as a
time-dependent harmonic oscillator.  This follows from the fact that the
translated action will possess the form
\begin{equation}
\label{161}
{\cal L } =
- q \dot{p} - \frac{p^2}{2m} - \frac{1}{2}
\frac{ \partial^2 V ( q_c ) }{ {\partial q_c}^2} q^2 - \ldots
\end{equation} where $q_c$ is a classical solution to the original equations
of motion consistent with the boundary conditions $q_c (t = 0) = q_i$ and
$p_c (t = T) = p_f$.  The presence of a set of well-defined eigenvalues of
the associated eigenvalue problem is of central importance in determining
tunneling rates and stability of states in the quantum theory and is
intimately related to Morse theory \cite{Morse}.

An canonical transformation approach to the remaining quadratic path
integral, effectively a time-dependent harmonic oscillator with the boundary
conditions $q_i = p_f = 0$, will be used to obtain an approximate
evaluation.  This begins by defining the time-dependent frequency $\omega (
t )$ by
\begin{equation}
\label{162}
( \omega ( t ) )^2 = \frac{1}{m} \frac{\partial^2 V( q_c ) }{ {\partial
q_c}^2 } \; .
\end{equation}
The right-hand side of (\ref{162}) can be negative for a wide variety of
circumstances.  For example, the potential $ V(q) = - \beta q^2 + \lambda
q^4$ gives rise to negative values for $\omega^2$ along any trajectory that
passes through the range of values $q^2 < \beta / 6 \lambda$.  As a result
many trajectories will generate an imaginary value for $\omega$ for intervals
of $t$.

The time-dependent canonical transformation to be used is given by
\begin{equation}
\label{163}
F = - \frac{p^2}{ 2 m \omega (t) } \tan Q \; ,
\end{equation}
where the time-dependent frequency of (\ref{162}) appears in (\ref{163}).
Suppressing all terms of $O(\Delta)$ and using result (\ref{90c}), the
transformed Hamiltonian for this case is given by
\begin{equation}
\label{164}
\tilde{H} = \omega(t) P + \frac{P}{2 \omega(t) } \frac{\partial \omega(t)}
{\partial t} \sin 2 Q \; .
\end{equation}
Clearly, suppressing the $O(\Delta)$ terms is not valid in this case since
the transformed Hamiltonian is no longer cyclic.  As a result, the analysis
that follows must be considered as an attempt at an approximate but
nonperturbative evaluation of the path integral.  Hamilton's equations of
motion are given by
\begin{eqnarray}
\label{165}
\dot{Q} & = & \omega (t) +
\frac{\partial \omega(t)}{\partial t}
\frac{\sin 2Q}{2 \omega(t) } \; , \\
\label{166}
\dot{P} & = & - \frac{\partial \omega(t)}{\partial t}
\frac{\cos 2Q}{ \omega(t) } P \; .
\end{eqnarray}
The solution to (\ref{165}) depends upon the form of $\omega(t)$, but in
general it cannot be formally expressed as an integral.  The solution can be
obtained by iteration or can be approximated.  To lowest order, the form for
$Q$ consistent with the boundary condition $q_i = 0$ is given by
\begin{equation}
\label{167}
Q(t) \, \approx \, \int_0^t d \tau \, \omega ( \tau ) \; .
\end{equation}
It is not difficult to see that (\ref{167}) is accurate for small values
of $t$, and hence for $T$ small.  Once the form for $Q(t)$ is known, it
is straightforward to solve (\ref{166}) by formal integration to obtain
\begin{equation}
\label{168}
\frac{P_f}{P_i} =  \exp \left \{ - \int_0^T {\rm dt} \, \left (
\frac{\partial \omega(t)}{\partial t}
\frac{\cos 2Q(t)}{ \omega(t) } \right ) \right \} \; .
\end{equation}
The classical action along the trajectory given by (\ref{165}) vanishes,
while by virtue of the boundary conditions, $F_i = F_f = 0$. The entire
translated path integral reduces to the prefactor generated by the
anomalies, and this is given by
\begin{equation}
\label{169}
\frac{1}{ \sqrt{2 \pi \hbar \cos Q(T) } }
\exp \left \{ - \int_0^T {\rm dt} \, \left (
\frac{\partial \omega(t)}{\partial t}
\frac{\cos 2Q(t)}{4 \omega(t) } \right ) \right \} \; .
\end{equation}
Result (\ref{169}) is, of course, dependent on the original form of the
interaction prior to translation as well as the values of $p_f$ and $q_i$.
This is because the functional form for $\omega(t)$ depends on the original
form of the interaction and the boundary conditions of the trajectory
through (\ref{162}).  Combining (\ref{169}) with the value of the original
action along the classical trajectory gives a new nonperturbative evaluation
of the original path integral.

\end{document}